\documentclass[letterpaper, 10 pt, conference]{ieeeconf}  

\IEEEoverridecommandlockouts                              

\usepackage{cite}
\usepackage{amsmath,amssymb,amsfonts}
\usepackage{algorithmic}
\usepackage{algorithm}
\usepackage{mathrsfs}
\usepackage{graphicx}
\usepackage{textcomp}
\usepackage{tabularx}
\usepackage{tabularray}
\usepackage{multirow}
\usepackage{xcolor}
\usepackage{diagbox}
\usepackage{booktabs}
\usepackage{subfigure}
\usepackage{tikz}
\usepackage{subcaption}
\usepackage{makecell,rotating}
\newcommand*\emptycirc[1][0.7ex]{\tikz\draw (0,0) circle (#1);} 
\newcommand*\halfcirc[1][0.7ex]{%
	\begin{tikzpicture}
	\draw[fill] (0,0)-- (90:#1) arc (90:270:#1) -- cycle ;
	\draw (0,0) circle (#1);
	\end{tikzpicture}}
\newcommand*\fullcirc[1][0.8ex]{\tikz\fill (0,0) circle (#1);} 
\newcommand*\halfsquare[1][1.4ex]{
	\begin{tikzpicture}
		\fill[black] (0,0) rectangle (#1/2,#1); 
		\draw (0,0) rectangle (#1,#1); 
	\end{tikzpicture}
}

\title{\LARGE \bf
Threats and Defenses in Federated Learning Life Cycle: A Comprehensive Survey and Challenges
}

\author{Yanli Li,  Zhongliang Guo, Nan Yang, Huaming Chen, Dong Yuan, and Weiping Ding*
\thanks{Yanli Li is with the School of Artificial Intelligence and Computer Science, Nantong University, Nantong 226019, China, and also with the School of Electrical and Computer Engineering, The University of Sydney, Sydney, NSW 2006, Australia (e-mail: yanli.li@sydney.edu.au).} 
\thanks{Zhongliang Guo is with the University of St Andrews, St Andrews, Fife, KY16 9SX, UK (e-mail: zg34@st-andrews.ac.uk).}
\thanks{Nan Yang, Huaming Chen, and Dong Yuan are with the School of Electrical and Computer Engineering, The University of Sydney Sydney, NSW 2006, Australia (e-mail: \{n.yang, huaming.chen, dong.yuan\}@sydney.edu.au,}
\thanks{Weiping Ding is with the School of Artificial Intelligence and Computer Science, Nantong University, Nantong 226019, China (e-mail: dwp9988@163.com).}
\thanks{{*}Corresponding author.}
}

\begin{document}

\maketitle
\thispagestyle{empty}
\pagestyle{empty}

\begin{abstract}
Federated Learning (FL) offers innovative solutions for privacy-preserving collaborative machine learning (ML). Despite its promising potential, FL is vulnerable to various attacks due to its distributed nature, affecting the entire life cycle of FL services. These threats can harm the model's utility or compromise participants' privacy, either directly or indirectly. In response, numerous defense frameworks have been proposed, demonstrating effectiveness in specific settings and scenarios. To provide a clear understanding of the current research landscape, this paper reviews the most representative and state-of-the-art threats and defense frameworks throughout the FL service life cycle. We start by identifying FL threats that harm utility and privacy, including those with potential or direct impacts. Then, we dive into the defense frameworks, analyze the relationship between threats and defenses, and compare the trade-offs among different defense strategies. Finally, we summarize current research bottlenecks and offer insights into future research directions to conclude this survey. We hope this survey sheds light on trustworthy FL research and contributes to the FL community.
\\ 

\textbf{\textit{Index Terms}— Federated Learning (FL), Adversarial Machine Learning (AML), Privacy, Robustness.}
\end{abstract}


\section{INTRODUCTION}
Nowadays, smart devices have become integral to people's daily lives. While engaging with these applications or smart services, users generate tens of billions of data points per second. This data, regarded as a valuable resource for fueling data-driven models, but remains underutilized due to the centralized nature of traditional training paradigms \cite{li2021survey, gabrielli2023survey}. In particular, centralized data collection necessitates stable network connections and high-speed broadband, conditions that often conflict with the realities of edge devices. On the other hand, transferring data from the user's domain raises significant privacy concerns, particularly when dealing with sensitive financial, medical, and social media information. As phrased by authors in \cite{kairouz2021advances}, “data remains isolated within individual owners' hands, creating data silos.”

To mitigate the problem of data silos, federated learning (FL) was introduced by Google in 2017 \cite{mcmahan2017communication}. This new machine learning (ML) paradigm allows participants to collaborate in a loosely federated manner while retaining a degree of “autonomy” \cite{kairouz2021advances}. Under the FL paradigm,  participants locally train their model and only upload model updates for aggregation. Due to its privacy-preserving characteristics, FL has attracted significant attention from both academia and industry, leading to the deployment of numerous real-world applications. Notable examples include Google's smart keyboard Gboard \cite{hard2018federated}, Apple's intelligent voice assistant Siri \cite{apple}, and WeBank's finance risk prediction service for reinsurance \cite{WeBank}. A typical FL service life cycle \cite{kairouz2021advances, mcmahan2017communication, li2021survey} begins with a service provider publishing a task and selecting participants. Each selected client then receives the global model, trains it locally, uploads the gradients, and participates in the aggregation process. Once the global model achieves the desired performance, the server distributes rewards and releases the application to the public.

Despite its promise, FL is vulnerable to various attacks due to its distributed nature \cite{li2024secure}. Based on the capacity of adversaries, these threats can completely  \cite{damaskinos2019aggregathor, el2022genuinely, fang2020local} or partially \cite{9806416, li2022backdoor, baghbani2022application} harm the utility of model, break the privacy of participants \cite{shokri2017membership, nasr2019comprehensive, zhang2023privacy}, and could happen at any phase during the FL service life cycle. In response, numerous defense frameworks \cite{ozdayi2021defending, taheri2020defending, cao2020fltrust} have been proposed and shown effectiveness across different settings and scenarios. Given the extensive security research, the FL community urgently needs comprehensive reviews to provide a clear understanding of the current research landscape. 

Although some reviews exist \cite{xia2023poisoning, wen2023survey, mothukuri2021survey}, their scope is generally limited to threats leading to direct, serious harm and around the local training phase, failing to provide a comprehensive picture of the research field. To mitigate the research gap, we present this review, which aims to comprehensively investigate all threats throughout the FL life cycle, along with their corresponding defenses. Unlike existing reviews, our study considers both direct and potential threats and expands the scope to encompass the entire FL life cycle. 

\begin{figure*} [h]
    \centering
    \includegraphics[width=1.0\linewidth]{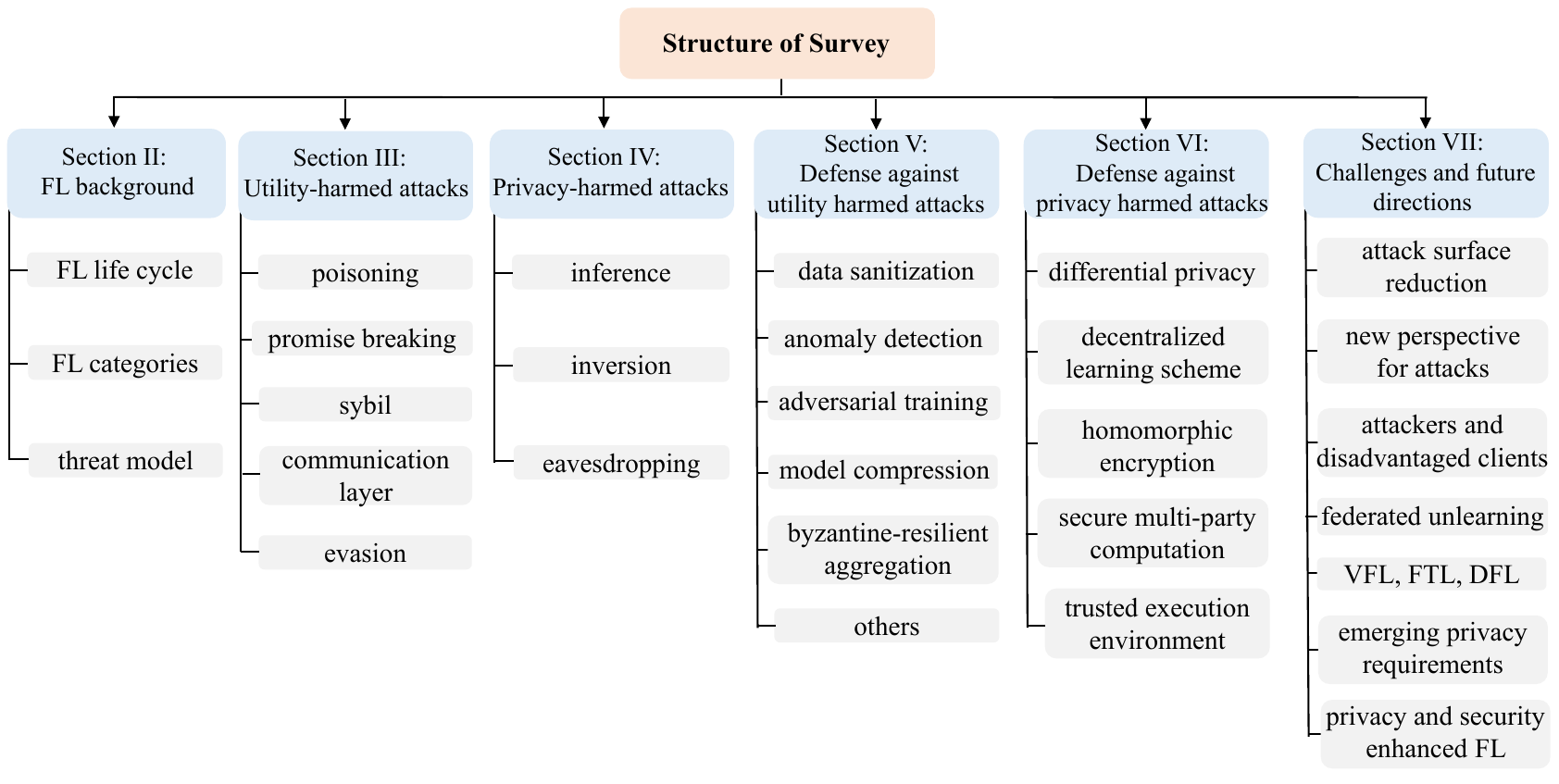}
    \caption{Organization of the survey.}
    \label{organization}
\end{figure*}

In summary, the main contributions of this survey are as follows:
\begin{itemize}
    \item Providing the entire life cycle of FL services, from the learning task publish to service delivery; identifying each learning phase's vulnerabilities, the potential threats' impact, and actors.  
    \item Identifying and comparing the utility-harmed threats in the FL environment, providing their attack and poisoning strategies, ranging from those with the potential to direct impact.
    \item Identifying and comparing the privacy-harmed threats in the FL environment, ranging from the application to the network layer.
    \item Classifying the existing defense frameworks, summarizing their effectiveness when defending against various utility-harmed or privacy-harmed threats, and discussing their assumptions and trade-offs.
    \item Discussing the existing bottlenecks of trustworthy FL, identifying the emerging requirements, and providing insights about future trends. 
\end{itemize}

The rest of this paper is organized as follows: 
We provide the background information of FL in Section \ref{SEC2}, including its life cycle, vulnerabilities, and categories. Sections \ref{SEC3} and \ref{SEC4} review the most representative and state-of-the-art threats targeting the utility and privacy of FL systems. In Sections \ref{SEC5} and \ref{SEC6}, we summarize the defense frameworks for addressing FL privacy concerns and enhancing FL robustness, compare their effectiveness, and discuss their trade-offs. Section \ref{SEC7} offers insights into future research trends, and Section \ref{SEC8} concludes the survey. For better readability, we present a diagram in Figure \ref{organization}, illustrating the organization of the survey.

\begin{figure*} [h]
    \centering
    \includegraphics[width=1\linewidth]{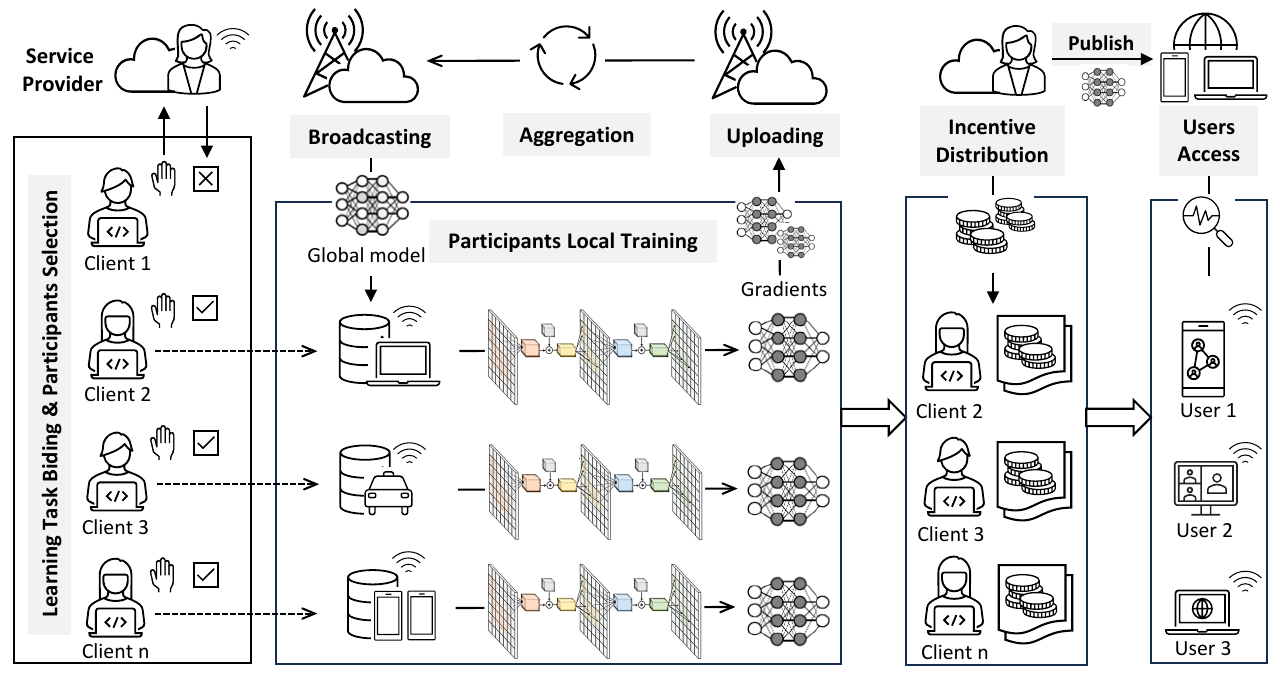}
    \caption{Federated learning life cycle.}
    \label{lifecycle}
\end{figure*}

\section{Federated Learning Background}\label{SEC2}
In this section, we introduce the FL service life cycle, different types of FL, and FL threat models to provide the survey's background.

\subsection{Federated Learning Life cycle}
\textbf{0. Learning task bidding and participants selection:} The life cycle of FL begins when service providers publish a learning task within FL communities or mention it in an application's user terms. Interested clients submit bids to participate based on the specified requirements and promised rewards. These bids include commitments of their computing resources, bandwidth, network capabilities, and training data. Service providers then review these commitments and select the participants for the learning task.

\textbf{1. Global model broadcasting:} The server, which can be the service provider or another third-party computing center, broadcasts the current global model to all selected participants. During the initialization of the learning task (the first learning round), the server also specifies the conditions for concluding both local and global learning iterations.

\textbf{2. Participants' local training:} Once the current global model is received, each participant trains the model locally using their private data. This training process continues for several rounds until preset conditions, such as a specified number of learning rounds or a desired model accuracy, are met. Subsequently, the participants send their local model updates back to the server, completing the local training phase.

\textbf{3. Aggregation and global model update:} The server aggregates the collected local model updates using a predefined aggregation rule. The resulting outcome serves as the gradient to update the global model. If the end condition has not been reached, the updated global model is then broadcast to all participants for the next round of training.

\textbf{4. Incentive distribution:} When the global model achieves the expected performance, the server stops broadcasting, and the training task is finalized. The service provider then evaluates each participant's contribution and distributes incentives accordingly, which could be either profit or non-profit.

\textbf{5. Model publish, and user access:} Finally, the service provider publishes the FL model or FL-based service online. Users can then access and utilize the application for various purposes, such as querying, making predictions, and more.

Figure \ref{lifecycle} illustrates the life cycle of FL service.

\subsection{Federated learning Categories}
The FL can be categorized based on the participants' data distribution or the system architecture \cite{Gao2022survey}.

\begin{figure} [h]
    \centering
    \includegraphics[width=1\linewidth]{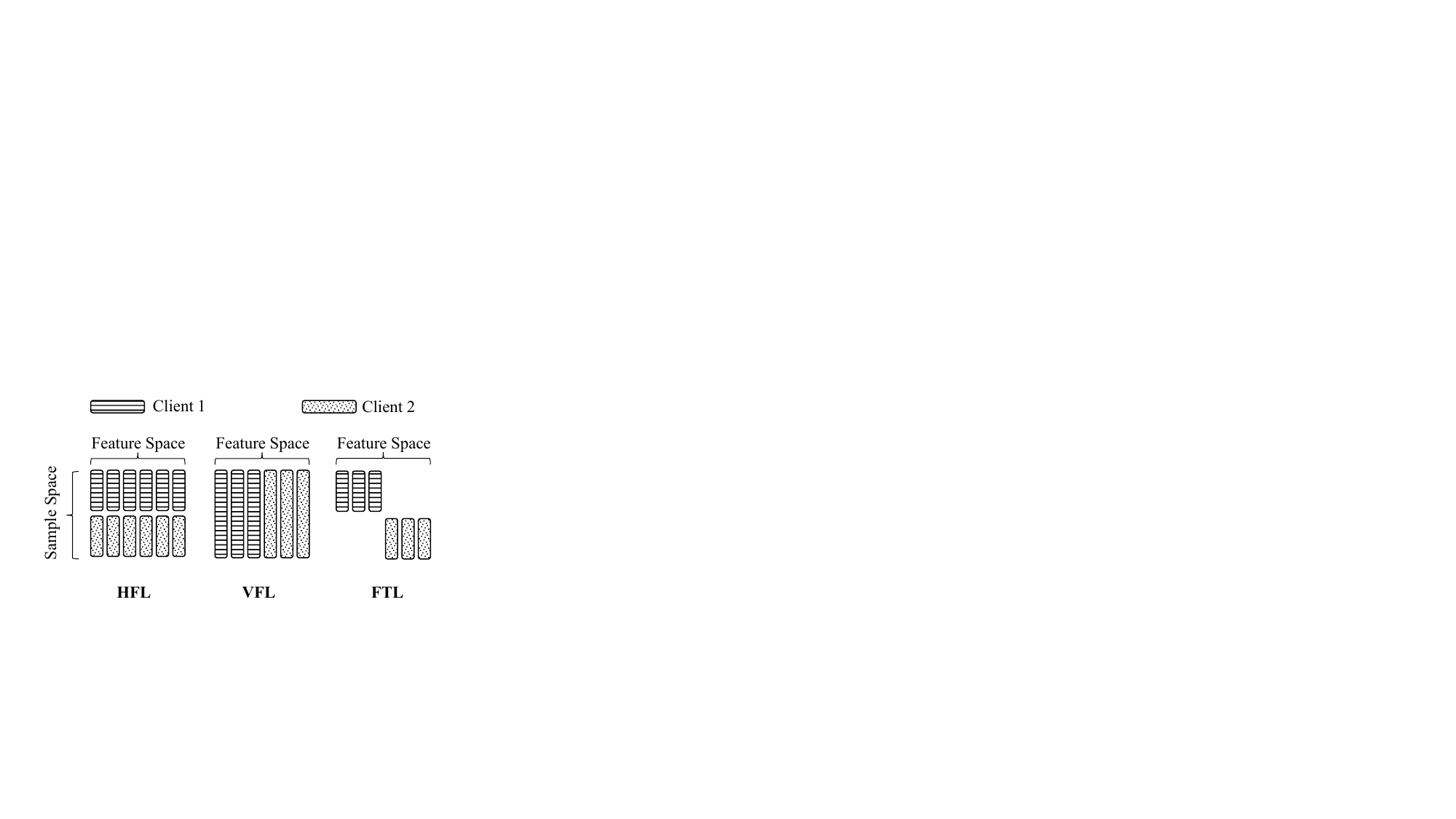}
    \caption{Illustration of the clients' data distribution of HFL, VFL, and FTL.}
    \label{HVTFL}
\end{figure}

Based on the distribution of data features and data samples, FL can be classified as horizontal federated learning (HFL), vertical federated learning (VFL), and federated transfer learning (FTL). Specifically, HFL is applicable to cases in which participants own different data samples but the same feature spaces. For example, in rare disease research, hospitals may have distinct patients but their records own highly similar features. Contrary to HFL, the VFL participants overlap in sample spaces but differ in feature spaces. For example, banks and telecom operators in a city may keep records of the same citizens, but these records contain different features. FTL integrates the concepts of FL and transfer learning, where transfer learning leverages knowledge acquired from one domain to enhance performance in another domain. FTL addresses the challenge of limited overlapping samples and features by training a model on a large public dataset.
Figure \ref{HVTFL} illustrates the clients' data distribution of HFL, VFL and FTL.

\begin{figure} [h]
    \centering
    \includegraphics[width=1\linewidth]{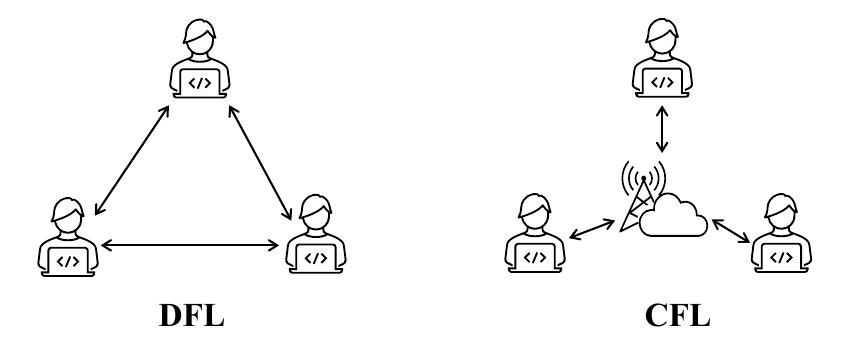}
    \caption{Illustration of CFL and DFL system architectures.}
    \label{CDFL}
\end{figure}

On the other hand, FL can be divided into centralized federated learning (CFL) and decentralized federated learning (DFL) based on the system architecture. Under CFL, a server organizes the whole learning process, including models broadcast, collection and aggregation. In contrast, DFL does not involve the central server. Following a preset rule, participants can directly exchange their models and automatically perform aggregation. Without further notification, we use FL to represent CFL in this survey.
Figure \ref{CDFL} illustrates the system architectures of CFL and DFL.

\begin{table*}[htbp]
\caption{Illustration of the vulnerabilities throughout the FL life cycle and the threat's impacts.}
\begin{center}
\begin{tabular}{c|c|c|c|c|c|c|c}
\toprule[1pt]
\midrule
\multicolumn{2}{c|}{\textbf{FL Life Cycle}}&\textbf{Client Selection} &\textbf{Model Broadcast}&\textbf{Local Training}&\textbf{Aggregation}&\textbf{Incentive}&\textbf{Users Access}\\
\midrule
\multicolumn{2}{c|}{Actor}&Server, Potential clients&Server&Clients&Sever&Server, Clients&Users\\
\multicolumn{2}{c|}{Interacted Surface}&Cloud, Ground&Cloud, Network&Ground, Network&Cloud&Cloud, Ground&Ground\\
\midrule
\multirow{5}{*}{\rothead{Utility}}&Poisoning  &\fullcirc &\fullcirc &\emptycirc &\fullcirc &\fullcirc&\fullcirc\\
&Promise Breaking  &\emptycirc \  \halfcirc &\fullcirc &\halfcirc &\fullcirc &\halfcirc&\fullcirc\\
&Sybil  &\emptycirc &\fullcirc &\emptycirc & \fullcirc&\fullcirc&\fullcirc\\
&Communication  &\fullcirc & \emptycirc \  \halfcirc & \halfcirc & \fullcirc&\fullcirc&\fullcirc\\
&Evasion  &\fullcirc & \fullcirc & \fullcirc & \fullcirc & \fullcirc &\halfcirc\\
\midrule
\multirow{3}{*}{\rothead{Privacy}}&Inference &$\blacksquare$& $\blacksquare$ &$\square$ & $\square$ &$\blacksquare$ & $\square$\\
&Inversion &$\blacksquare$ & $\blacksquare$ &$\blacksquare$ & $ \square$ $\halfsquare$ & $\blacksquare$&$\blacksquare$\\
&Eavesdropping&$\blacksquare$& $\square$ &$\square$ & $\blacksquare$ &$\blacksquare$ & $\blacksquare$\\
\midrule
\multicolumn{8}{c}{{\fullcirc}\ $\blacksquare$ Non-exist, \ \  \halfcirc \ $\halfsquare$ \ Exist and Potential impact, \ \  \emptycirc \ $\square$  \ Exist and Direct impact}\\
\bottomrule[1pt]
\end{tabular}
\label{tab1}
\end{center}
\end{table*}

\subsection{Threat Models in Federated Learning Life Cycle}
In FL, each participant has a certain degree of autonomy and can independently conduct local training. While the federated architecture potentially enhances participants' privacy, it also introduces new attack surfaces. Beyond the training phase attacks that directly degrade FL model accuracy and compromise client privacy, dishonest behavior in other phases of the FL life cycle can also negatively impact the model. For instance, dishonest self-reporting during participant selection can lead to insufficient computing or data resources during training, resulting in lower-than-expected model performance. In light of these scenarios, we propose a broader definition of threats and attackers in this survey:\\

\textit{\textbf{Threats (i.e., attacks)} are the misbehavior that may directly or potentially harm FL models' performance (i.e., utility) or privacy.}

\textit{\textbf{Attackers} are the actors who intentionally launch threats.}\\

\textbf{Vulnerabilities in FL:} Today, numerous studies have demonstrated that attacks can occur at any stage throughout the entire FL life cycle. Initially, a dishonest client may exaggerate its capabilities to secure a spot in the selection process and join the learning task. Under the local training, the malicious clients may use modified training data to train the model, or directly perturb the trained model (or gradients) to degrade, even fully break the model performance. During server-client communication, the information transmitted can be replaced or tampered with by a "middleman." Once the training is complete, the dishonest participants can forge their consumption to defraud high returns. When the FL-based application is published, an attacker can use a crafted data item to mislead the application to generate unexpected outcomes. Although the server is assumed to be trustworthy in most cases, the server is potentially curious or hijacked during the training process. In those cases, the server may act maliciously and infer sensitive information from individual updates collected. We summarize and provide detailed discussions of the utility-harmed and privacy-harmed threats in Section \ref{SEC3} and \ref{SEC4}. Table \ref{tab1} illustrates the vulnerabilities throughout the FL life cycle and the threat's impacts.

\textbf{Attacker’s Knowledge and Capacity:} Attacker’s knowledge represents how much information of the FL system can be inspected by the adversary. From the perspective of model parameters, the attacker's knowledge can be categorized as black-box, grey-box, or white-box \cite{guo2024white, wen2023survey}. These three settings correspond to scenarios where the adversary has no access to the FL model, has the ability to inspect the model of a particular state (possibly a historical version), and can inspect the FL model parameters of the current learning round, respectively. Considering participants receive the global model to perform local training each iteration, the grey box is the most general setting in adversarial FL contents. As some FL algorithms introduce the byzantine-robust aggregation rule as a defense mechanism, the attacker’s knowledge is also considered as to whether the aggregation rule is exposed and reviewed by adversaries. 

The attacker's capacity in the FL context generally refers to the level of collusion among adversaries. Malicious clients are often assumed to constitute a certain percentage of the overall participants, and these attackers can collude to agree on the same attack strategy to maximize attack efficiency. In a non-collusion setting, the adversaries are considered to have different attacking objectives or strategies, resulting in a weaker attack effect. We note that some recent works \cite{kumar2023impact} classifies the attacker's capacity as passive or active, depending on whether the adversary can directly interfere with the global model or simply inspect it. However, we believe that the terms passive and active do not accurately reflect the capacity of the attackers but rather the characteristics of different phases in the FL life cycle. Specifically, participants can arbitrarily modify their local model during the training phase, while they can only access or visit the FL application during the pre- and post-training phases. Thus, in this survey, we do not discuss the passive or active nature of adversaries from their perspective. Instead, we emphasize the interaction methods of various parties to the FL system at different phases in the FL life cycle.

In some attack strategies, attackers may need to perform large-scale grid searches or complex computations to generate the desired crafted parameters. Under these scenarios, the attacker's capacity extends to computing resources, which could be at the average level of the overall clients or potentially unlimited.

\begin{figure*} [h]
    \centering
    \includegraphics[width=1.0\linewidth]{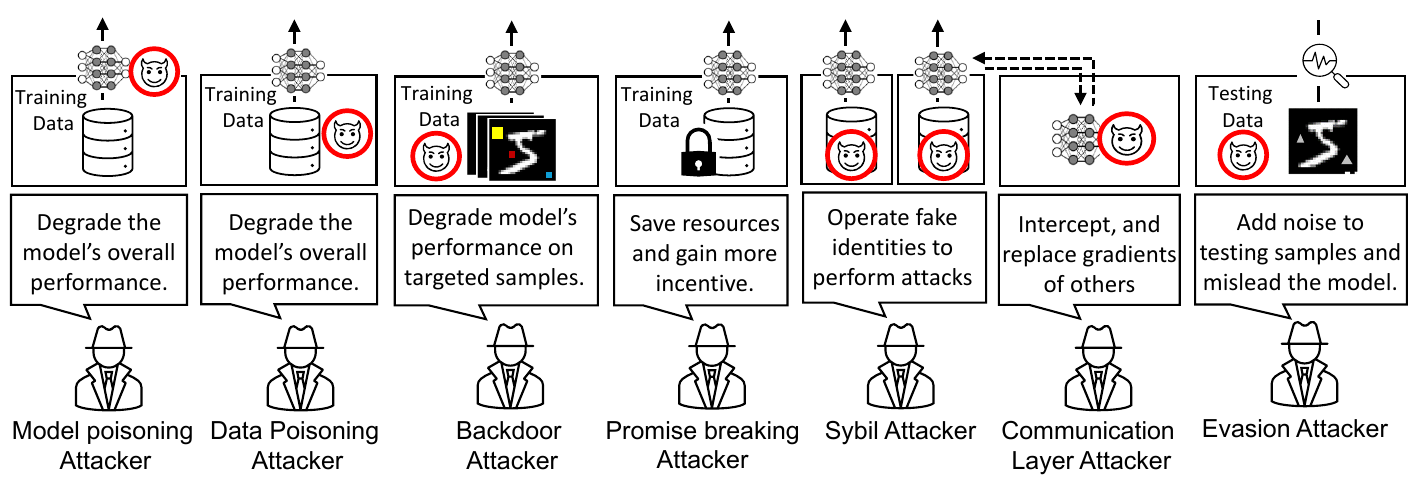}
    \caption{Demonstration and comparison of different FL utility-harmed attacks.}
    \label{compareu}
\end{figure*}

\section{Utility-harmed Attacks} \label{SEC3}
Compared to data-centralized collection learning schemes, FL provides a different way of training models collaboratively \cite{mcmahan2017communication, kairouz2021advances}. Instead of relying on a visible and centralized collection dataset, FL trains the model across a potentially vast array of unreliable devices, each equipped with private, uninspectable datasets. Furthermore, the local training process also introduces a new attack surface and brings opportunities for potential attacks \cite{zhou2021deep, fang2020local, 10121613, xiao2012adversarial}. Specifically, as data remains on the devices and only model updates are shared, adversaries could attempt to exploit this setup to compromise the model's utility or performance. We demonstrate and compare different FL utility-harmed attacks in Figure \ref{compareu}.

\subsection{Poisoning Attacks}
As one of the most representative adversarial attacks, poisoning attacks attempt to degrade the model performance or totally destroy the global model. Depending on the stage at which the poisoning occurs, poisoning attacks can be further categorized into model poisoning attacks and data poisoning attacks.

\subsubsection{Model poisoning Attacks}
Model poisoning attacks \cite{LI2024103936} refer to adversaries directly manipulating reports (local model updates) submitted to the service provider. One of the most straightforward strategies for executing model attacks involves introducing a fixed perturbation or substituting the benign gradient parameters with fixed values. For instance, Reverse Attack \cite{damaskinos2019aggregathor}, and Random Attack \cite{el2022genuinely} generate the malicious gradient by reversing and randomly replacing the original gradients. Under the Partial Drop Attack \cite{el2022genuinely}, the adversary replaces a preset percentage of the benign gradient parameters as 0; the variants of Partial Drop Attack enlarge the attack effectiveness by using other malicious values if the benign gradients naturally carry a large amount of 0. Although these attack strategies can effectively degrade the global model performance and decrease the testing accuracy, the malicious gradients are unavoidably away from the benign gradients, which can be identified and discarded by the byzantine resilient FL methods. 

To enhance the attacking performance and avoid being identified by the defender, some existing attacks \cite{zhou2021deep, 10121613, fang2020local} design more dynamic attack strategies. For instance, Little is Enough Attack \cite{baruch2019little} first normally trains the model through the data of the clients controlled. Then, the adversary statistics the mean value and standard deviation (Std) of the gradients owned. Based on the statistics, the adversary subsequently generates the crafted gradient by adding a scaled Std on the mean value. The Fall of Empires Attack \cite{xie2020fall} follows a similar strategy, which generates the malicious gradient by timing the mean value with a preset scalier. To maintain the stealth of model poisoning attack, research \cite{zhou2021deep} proposes an optimization-based model poisoning attack, which only manipulates a small fraction of the local model updates instead of all gradient parameters. Under the study \cite{zhou2021deep}, adversary training is organized into two tasks: the main task and the adversarial task. In the main task, the adversary trains the model in the usual manner. Conversely, in the adversarial task, the adversary embeds adversarial features into the neural network's redundant space to enhance the attack's persistence.

Recently, Local Model Poisoning Attack \cite{fang2020local} proposes a more flexible attacking algorithm that adopts the attack strategy for different defense frameworks. Based on the information owned (partial or full knowledge), this attack first infers the optimal global model updates and subsequently formulates the optimization problem that aims to reverse the global model update the most along the infers. This work \cite{fang2020local} applies the attack in Krum \cite{blanchard2017machine}, Trimmed-mean \cite{yin2018byzantine}, and Median \cite{yin2018byzantine} frameworks and consequently generates the Krum Attack and Trim Attack. The study \cite{10121613} targets the Krum framework with an emphasis on evasion detection, specifically through covert model poisoning (CMP). Research \cite{10121613} first formulates the model poisoning as an optimization problem, aiming to minimize the Euclidean distance between the manipulated model and a designated one, all within the constraints set by Krum. Based on the solution to the optimization problem, \cite{10121613} subsequently formulates CMP algorithms to counter the Krum framework. To improve the practicability, a low-complexity CMP algorithm is introduced and achieves the optimization complexity reduction. We summarize various model poisoning strategies in Table \ref{tab2}.

\begin{table}[htbp]
\caption{Illustration of various model poisoning strategies.}
\begin{center}
\begin{tabular}{|p{3.8cm}|p{3.8cm}|}
\hline
\multicolumn{2}{|c|}{\textbf{Nomination}}\\
\hline
{$G'$ \hfill Crafted gradients} & {$\dot{G}$ \hfill Gradients of benign clients}\\
{$\ddot{G}$ \hfill  Malicious gradients}&{$G$ \hfill Gradients of all clients}\\
{$M$ \hfill  Support matrix}&{$P$ \hfill Preset attack possibility}\\
{$z, \gamma$ \hfill  Attack scalar}&{$\mu$ \hfill Mean of gradients}\\
{$\sigma$ \hfill  Standard deviation}&{$\nabla$ \hfill Additive perturbation}\\
\hline
\end{tabular}

\vspace{0.5cm}

\begin{tabular}{c|c}
\toprule[1pt]
\midrule
\textbf{Attacks}&\textbf{Poisoning Strategy}\\
\midrule
Reverse \cite{damaskinos2019aggregathor}&$G'=-\ddot{G}$\\
\midrule
Mimic \cite{karimireddy2021byzantine}&$G'=\ddot{G_i}$\\
\midrule
Partial Drop \cite{el2022genuinely}&$G'=\ddot{G}\cdot M, M\sim Bernoulli(P)$\\
\midrule
Random \cite{el2022genuinely}&$G'=\ddot{G}\cdot M, M\sim Random(P)$\\
\midrule
LIE \cite{baruch2019little}&$G'=\mu-z\sigma$ or $\mu+z\sigma$\\
\midrule
FOE \cite{xie2020fall}&$G'=-z\mu$ or $z\mu$\\
\midrule
LMP \cite{fang2020local} partial &$G'\in$ \((\mu+3\sigma, \mu+4\sigma)\) or \((\mu-4\sigma, \mu-3\sigma)\)\\
\midrule
LMP \cite{fang2020local} full &$G'\in$ \((G_{max}, z\cdot G_{max})\) or  \((z\cdot G_{min}, G_{min})\)\\
\midrule
Min-Max \cite{shejwalkar2021manipulating} &$G'=f_{avg}(\dot{G})+\gamma \nabla ,$\\
 &$\mathop{\arg \max}\limits_{\gamma} max||G-\dot{G_i}||_2\leq max||\dot{G_i}-\dot{G_j}||_2$\\
\midrule
Min-Sum \cite{shejwalkar2021manipulating} &$G'=f_{avg}(\dot{G})+\gamma \nabla ,$\\
 &$\mathop{\arg \max}\limits_{\gamma} \sum||G-\dot{G_i}||_2^2\leq \sum||\dot{G_i}-\dot{G_j}||_2^2$\\
\midrule
\bottomrule[1pt]
\end{tabular}
\label{tab2}
\end{center}
\end{table}

\subsubsection{Data poisoning Attacks}
Different from model poisoning attacks that directly introduce the perturbation to the gradient, data poisoning attacks generate crafted gradients by training models based on poisoned data. Label Flipping Attack is one of the most representative data poisoning attacks. In Label Flipping Attacks, the adversary mismatches the training labels and training data to generate the malicious gradient, aiming to degrade the attacked class learning performance of the joint model. Study \cite{xiao2012adversarial} considers the support vector machine (SVM) algorithm and aims to find a label-flipping combination under a given budget so that a classifier trained on such data will have maximal classification error. Research \cite{tolpegin2020data} introduces the Label Flipping Attack in the deep learning scenarios and evaluates the key conditions for successfully performing attacks, including sufficient adversarial participants and continuous participation. To enhance the flexibility of the attacks, study \cite{lewis2023attacks} proposes the On-Off Label Flipping strategy. Under the On-Off Label Flipping strategy, the adversary first acts as a benign client for a period of time and builds a positive expectation in the defense system to strengthen the impact of the following updates. Then, the adversary betrays the system and performs an attack for a subsequent period of time. As the betrayal behavior lowers the reputation, the adversary may toggle back to acting as the benign client again after the One-Off attack and recover its reputation.

Label Flipping Attacks also show significant effectiveness in real-world applications. For instance, Study \cite{taheri2020defending} implements this attack within a malware detection task for Android platforms, employing a Silhouette Clustering-based Label Flipping Attack (SCLFA). This method involves attackers calculating the silhouette scores of data samples and targeting those with negative values for generating polluted data through label flipping. Similarly, research \cite{zhang2021label} illustrates the potency of Label Flipping Attacks in compromising spam filtering systems. By executing entropy-based flipping attacks, their study managed to elevate the false negative rate of Naive Bayes classifiers in the presence of label noise without impairing the classification of legitimate emails. Furthermore, study \cite{sharma2022catboost} explores the impact of Label Flipping Attacks on hardware Trojan detection systems, demonstrating how model performance could be substantially degraded through a stochastic hill-climbing search-based flipping strategy, incurring minimal costs for label manipulation. These instances underscore the critical need for robust defenses against Label Flipping Attacks in diverse application areas.

\subsubsection{Backdoor Attacks}
Backdoor Attack is a special type of poisoning attack, which aims to lead the global model to misbehave only on a selected minority of examples while maintaining good overall accuracy on all other examples \cite{9806416, li2022backdoor}. Recent studies indicate that the Backdoor Attack can be executed either by poisoning the clean training data or by introducing crafty perturbations to benign gradients. In other words, the Backdoor Attack can be categorized as either a model poisoning attack or a data poisoning attack based on the poisoning strategy. 

When performing Backdoor Attacks through the data poisoning strategy, the adversary should first decide on a trigger (backdoor) and consequently assign a selected label for the trigger carrier to form the poisoning data. This backdoor could be semantic or artificial. For semantic backdoor, study \cite{baghbani2022application} leverages the samples with uncommon attributes as the backdoor, such as the cars with unconventional colors (e.g., green), scenes containing a unique object (e.g., a striped pattern), or trigger sentences in word prediction problems that conclude with an attacker-selected target word. Similarly, research \cite{nguyen2020poisoning} proposes an attack strategy to introduce a backdoor in the FL-based IoT intrusion detection system. In this scenario, the adversary targets packet sequences originating from specific malware-driven malicious traffic. Different from the semantic backdoor that replies on existing share properties, the artificial backdoor manually poisons benign samples by introducing triggers. Study \cite{li2021survey} demonstrates the adversary can manually add pattern “L” at the images' corner to activate the backdoor, where the pattern “L” does not naturally exist in the benign samples. A recent study \cite{bagdasaryan2020backdoor} extends the scope of backdoor attacks to real-world settings. The research reveals that beyond digital triggers, physical triggers such as sunglasses, tattoos, and earrings can also serve as triggers for organizing backdoor attacks. 

Within the Vanilla FL framework, the most straightforward method to execute a backdoor attack using a model poisoning strategy is scaling the malicious updates containing backdoor information to dominate updates from benign clients. Study \cite{bagdasaryan2020backdoor} first employs the replace method, in which the attacker seeks to substitute the new global model with a poisoned one by a wisely chosen factor. While the scaling-replacement method proves effective in averaging aggregation, straightforward scaling appears to be naive to success under clipping and restricting defenses. To enhance the stealth of model poisoning attacks, research \cite{wang2020attack} proposes a projected gradient descent (PGD) attack strategy, which projects the crafted model on a small ball centered around the global model from the previous iteration. 

\subsection{Promise breaking}
During the FL life cycle, clients may pledge to provide their training data or computing resources to gain participation opportunities or claim their contributions to receive incentives. However, malicious clients can exploit the FL system to gain advantages, such as access to the joint model and incentives, without fulfilling their promises \cite{wan2021shielding, fraboni2021free}. Although these attackers do not deliberately poison the model, promise-breaking behaviors can lead to insufficient computing resources and training data, as well as unfair incentive distribution. This, in turn, may degrade learning performance and diminish participant motivation.

Research \cite{fraboni2021free} introduces a Free-riding Attack framework based on the Vanilla FL and shows that the adversary can perform a Free-riding Attack in each learning iteration. In particular, the Free-riding attacker does not perform local training but adds noise to the constructed parameter updates and applies Stochastic Gradient Descent (SGD) to maximize similarity with updates from other benign clients. Research \cite{wan2021shielding} proposes a novel Free-riding Attack in which the model is trained locally using a small dataset masqueraded as a large dataset to obtain more incentive rewards. Study \cite{lin2019free} introduces a Free-riding attack strategy, namely Random Weights Attack. Under this attack, the adversary creates a gradient update matrix with dimensions matching the received global model by randomly sampling each parameter from a uniform distribution within a pre-designed range.

Studies \cite{liu2020incentives, zeng2021comprehensive, weng2019deepchain} investigate dishonest reporting during intensive distribution. To cheat the reward system, malicious participants may falsely report their resource consumption or claim to have trained the joint model with high-quality data while actually using low-quality data, such as spam.

Table \ref{tab3} compares different promise-breaking threats.
\begin{table}[htbp]
\caption{Illustration of different promise-breaking attacks.}
\begin{center}
\begin{tabular}{c|c|c}
\toprule[1pt]
\midrule
\textbf{Promise Breaking}&\textbf{Phase}&\textbf{Objective}\\
\midrule
Free-riding  \cite{fraboni2021free, wan2021shielding, lin2019free}&Local training&Model acquisition\\
\midrule
{Dishonest reporting} & Client selection& Model acquisition \\
\cite{liu2020incentives, zeng2021comprehensive, weng2019deepchain} &Incentive distribution& Incentive / Reward\\
\midrule
\bottomrule[1pt]
\end{tabular}
\label{tab3}
\end{center}
\end{table}

\subsection{Sybil Attacks}
The Sybil Attack refers to a single attacker (or a small amount attackers) joining the system with multiple colluding identities to enhance the stealth and effectiveness of an adversarial attack. Given that some traditional FL permits selected participants to freely join and exit during training tasks, it becomes inherently susceptible to Sybil Attacks. 

The research \cite{fung2018mitigating} first introduces Sybil Attacks within the FL context. It demonstrates Sybil Attacks can significantly enhance the impact of other attacks (such as the Label Flipping Attack) by misleading the global model into classifying handwritten digits “1” as “7” through only two sybils. Study \cite{fung2020limitations} builds upon the findings of \cite{fung2018mitigating}, delving deeper into the vulnerability of FL to Sybil Attacks, as well as exploring associated attack strategies and objectives. This work \cite{fung2020limitations} categorizes and names the Denial-of-Service (DoS) attacks instigated by Sybil Attacks as Training Inflation and demonstrates Sybil Attacks can be performed from various FL dimensions. From the data distribution perspective, Sybils can use identical datasets, different datasets, or synthetic data to train local models and generate crafted updates \cite{LI2024120475}. On the other hand, to dominate the benign clients and amplify their influence on the system, Sybils might coordinate amongst themselves before producing the subsequent series of model updates, achieving different levels of coordination.

\subsection{Communication Layer Attacks}
FL continuously updates the joint model, which relies on communications between the server and multiple participants. However, the large volume of data exchange and frequent communication drives a significant communication bottleneck and brings new attack interfaces for adversaries. 

The man-in-the-middle (MiTM) attacks \cite{agrawal2018detection}, as one of the most representative network layer attacks, has been introduced in FL by recent studies \cite{gabrielli2023survey, 9994772}. As FL leverages the single server to aggregate model updates and organize the learning process, performing MiTM between the server and participants can effectively block the server and lead to a single point of failure. Research \cite{yao2018two} explores attacking the FL communication bottleneck by decreasing the bandwidth, delaying the response, and increasing the struggle possibility. Through performing attacks on the communication of FL, \cite{yao2018two} shows the model convergence and performance can be degraded significantly.
\section{Privacy-harmed Attacks} \label{SEC4}
\begin{figure*} [h]
    \centering
    \includegraphics[width=1.0\linewidth]{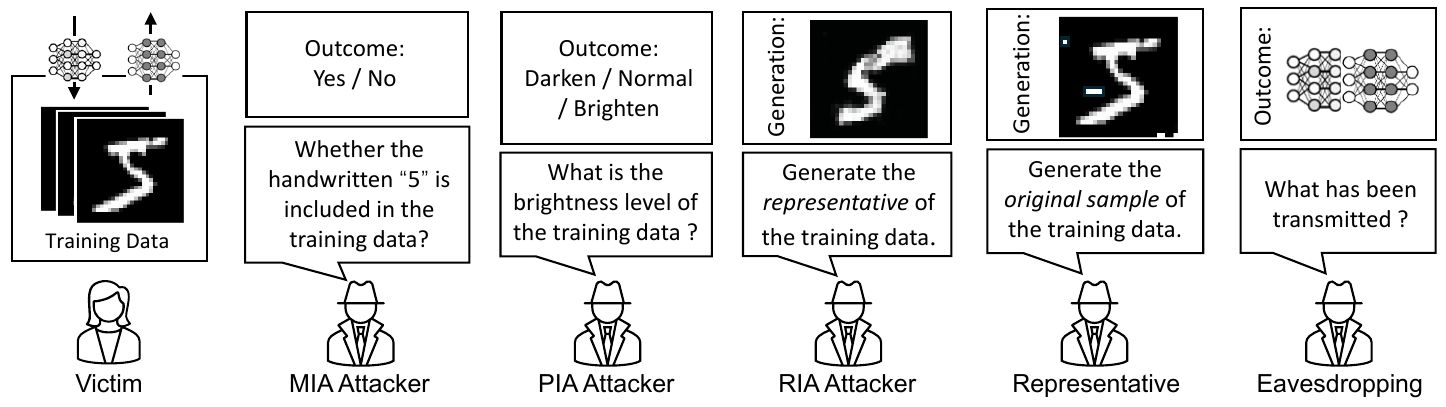}
    \caption{Demonstration and comparison of different FL privacy-harmed attacks.}
    \label{compare}
\end{figure*}
\subsection{Evasion Attacks}
In evasion attacks, the adversary deliberately adds slight malicious perturbations to input samples during the inference phase, causing the classifier to misclassify the sample with a high probability. One of the most representative works of the evasion attack has been introduced by \cite{goodfellow2014explaining}, which is a so-called adversarial example. By adding unnoticeable pixel perturbation on a panda image, \cite{goodfellow2014explaining} successfully changes the classification of GoogLeNet \cite{szegedy2015going} from panda to gibbon with 99.3\% confidence.

Based on the information owned by the adversary, the evasion attacks could be further divided into white-box and black-box attacks. In the white-box setting, the adversary has full knowledge of the learning algorithm and model parameters and can craft adversarial examples based on the information. For instance, studies \cite{madry2018towards, kurakin2017adversarial} generate adversarial examples by maximizing the loss function, subject to a norm constraint, using constrained optimization techniques like projected gradient ascent. On the other hand, the black-box setting refers to the attackers having no knowledge about the algorithm and parameter information of the FL system and generating adversarial samples through the interaction process or query access with the system. Considering the black-box setting, \cite{chen2017zoo} proposes a zeroth order optimization (ZOO) based attack to estimate the target model updates and generate adversarial examples consequently. Similarly, Boundary Attack \cite{brendel2018decision} has been designed in the same scenario. When performing Boundary Attack \cite{brendel2018decision}, the adversary first identifies the boundary separating adversarial and non-adversarial samples, then subtly introduces perturbations along this boundary to produce adversarial examples.

A recent work \cite{pang2022attacking} introduces an Evasion Attack in the Vertical FL (VFL) scenarios and proposes the Adversarial Dominating Inputs Attack. Unlike the aforementioned adversarial samples that domain the whole feature space, the Adversarial Dominating Inputs Attack \cite{pang2022attacking} can dominate the inputs of other clients by controlling a fraction of the feature inputs and lead to the misclassification of specific inputs. In the meanwhile, the Adversarial Dominating Inputs Attack can also result in reduced and unfair incentive rewards for impacted participants due to their decreasing contributions.

While FL is designed to protect the privacy of participants, recent studies have shown that privacy-harmed attacks still exist in FL systems. These attack objectives include identifying members (MIA) or properties (PIA) of victim training data, generating representative or (RIA) original samples (Inversion Attacks) of victim training data, and eavesdropping on transmissions between the victim and the server. We demonstrate and compare different FL privacy-harmed attacks in Figure \ref{compare}.

\subsection{Inference Attacks}
During the training process of FL, the model will inadvertently learn the latent information of private data. Therefore, an honest but curious server or any other user can launch an inference attack to recover the original training data without any prior knowledge. Inference Attacks include model extraction attacks, attribute inference attacks, and membership inference attacks.
\begin{figure} [h]
    \centering
    \subfigure[Ground truth - ImageNet1K \cite{deng2009imagenet}]{
		\begin{minipage}[t]{1.0\linewidth}
			\centering
			\includegraphics[width=1.0\linewidth]{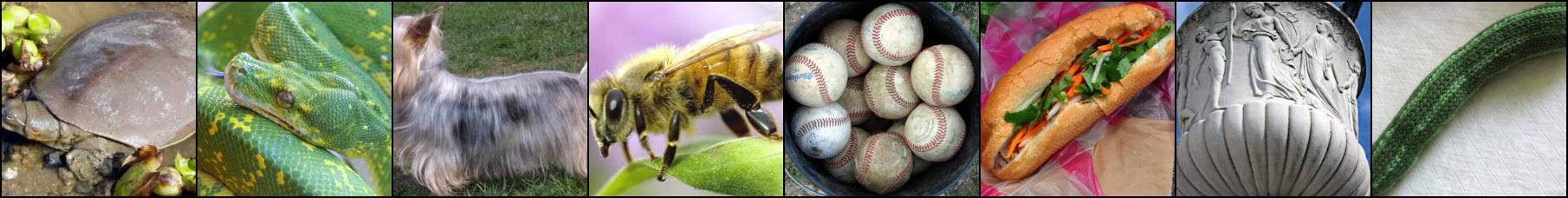}
		\end{minipage}%
        \label{mnist_flip26}
	}%
\newline
	\subfigure[Deep Leakage Gradient - NeurIPS'19\cite{zhu2019deep}]{
		\begin{minipage}[t]{1.0\linewidth}
			\centering
			\includegraphics[width=1.0\linewidth]{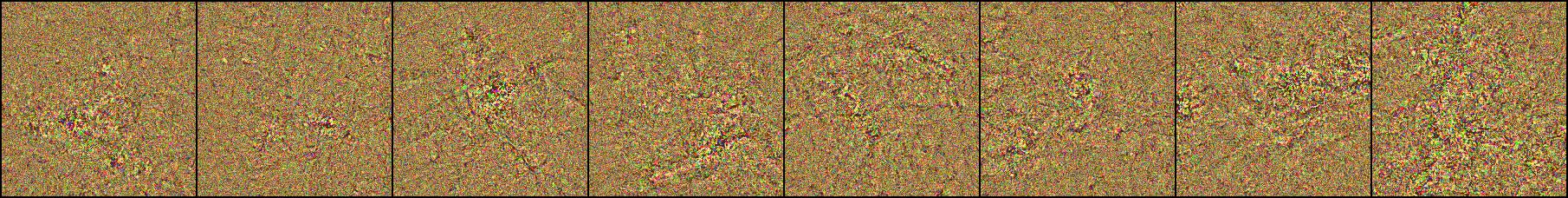}
		\end{minipage}%
        \label{mnist_flip58}
	}%
 \newline
    \subfigure[Inverting Gradients - NeurIPS’20 \cite{geiping2020inverting}]{
		\begin{minipage}[t]{1.0\linewidth}
			\centering
			\includegraphics[width=1.0\linewidth]{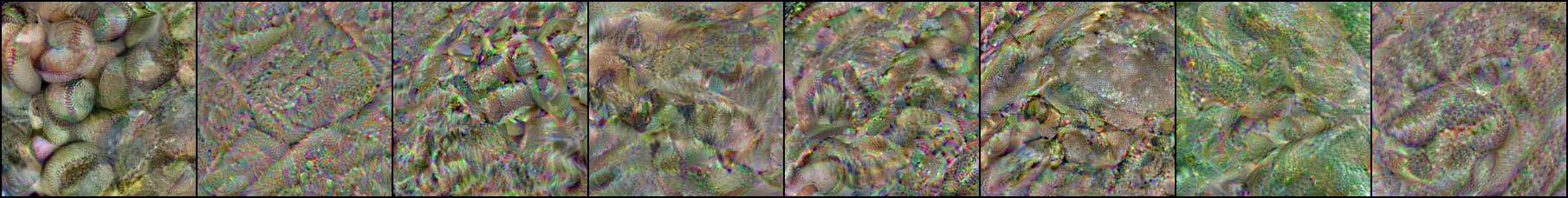}\\
		\end{minipage}%
        \label{mnist_flip58}
	}%
  \newline
    \subfigure[DeepInversion - CVPR’20 \cite{yin2020dreaming}]{
		\begin{minipage}[t]{1.0\linewidth}
			\centering
			\includegraphics[width=1.0\linewidth]{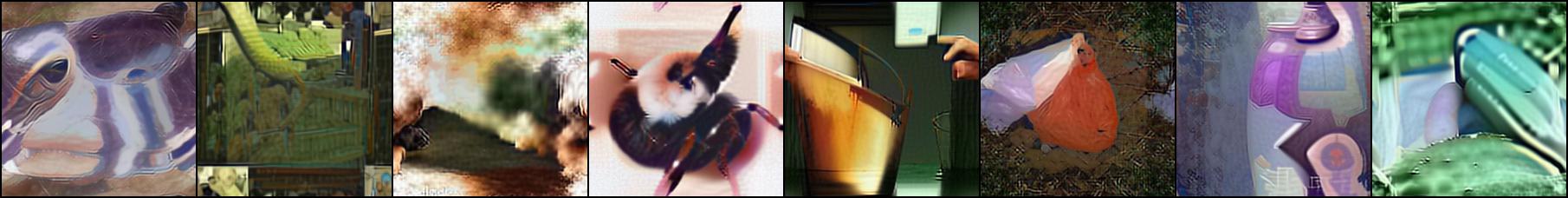}\\
		\end{minipage}%
        \label{mnist_flip58}
	}%
     \newline
    \subfigure[GradInversion - CVPR’21 \cite{yin2021see}]{
		\begin{minipage}[t]{1.0\linewidth}
			\centering
			\includegraphics[width=1.0\linewidth]{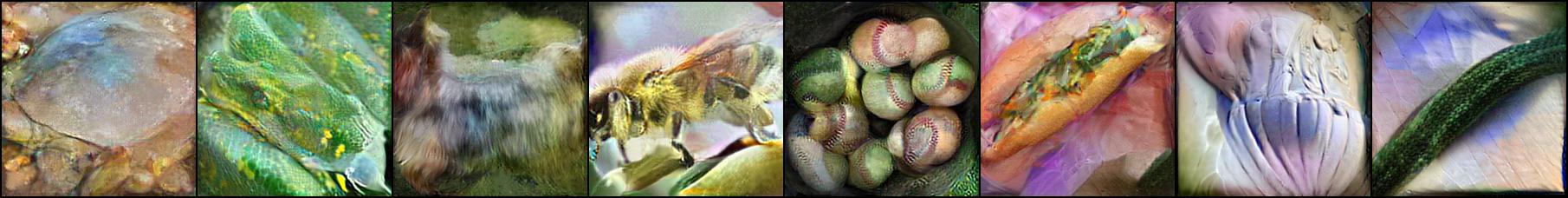}\\
		\end{minipage}%
        \label{mnist_flip58}
	}%
 \newline
     \subfigure[Generative Gradient Leakage - CVPR’22 \cite{Li_2022_CVPR}]{
		\begin{minipage}[t]{1.0\linewidth}
			\centering
			\includegraphics[width=1.0\linewidth]{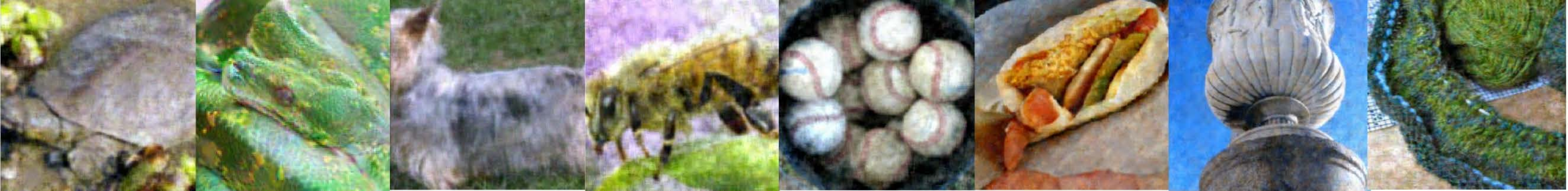}\\
		\end{minipage}%
        \label{mnist_flip58}
	}%
 \newline
    \subfigure[GradViT - CVPR’22 \cite{hatamizadeh2022gradvit}]{
		\begin{minipage}[t]{1.0\linewidth}
			\centering
			\includegraphics[width=1.0\linewidth]{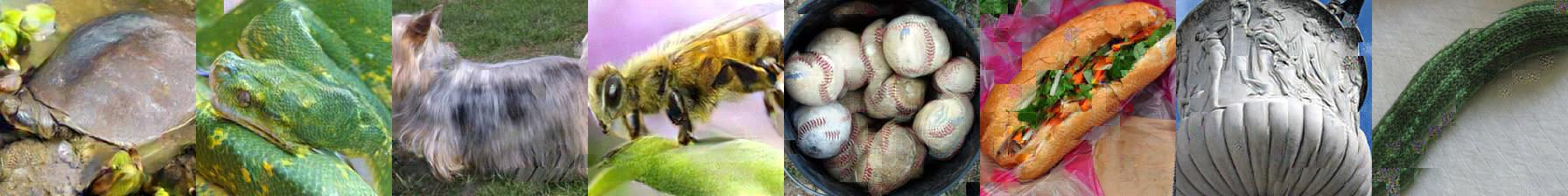}\\
		\end{minipage}%
        \label{mnist_flip58}
	}%
    \caption{The comparison of ImageNet batch gradient inversion outcomes under different inversion attacks.}
    \label{compare2}
\end{figure}

\subsubsection{Membership Inference Attack (MIA)}
Similar to MIAs in traditional machine learning scenarios, malicious clients or users can execute MIAs in FL applications to infer whether a data sample belongs to its training dataset. Study \cite{shokri2017membership} focuses on black-box within Machine Learning as a Service (MLaaS) settings. By accessing multiple generated shadow models, the study achieves high accuracy in inferring membership. ML-Leaks \cite{salem2019ml} reduces the cost of launching attacks and demonstrates the effectiveness of MIA using only a single (even without) shadow model. Study \cite{nasr2019comprehensive} considers the white-box settings and proposes the gradient ascent MIA strategy to magnify the presence of data points in others’ training sets. It demonstrates that all algorithms trained using Stochastic Gradient Descent (SGD) are susceptible to privacy vulnerabilities from MIAs. The study \cite{li2021membership} further extends MIAs to NLP learning tasks and effectively infers whether a specific text segment belongs to the training dataset.

\subsubsection{Property Inference Attacks (PIA)}
Property inference attacks aim to infer private features or characteristics (e.g., value, distribution, etc.) of the training data without direct exposure to the data. Although some surveys \cite{Gao2022survey, Gu2023Survey} categorize threats aimed at reconstructing training data as PIAs, we consider PIAs to primarily focus on inferring properties of the training data, rather than the data itself. Therefore, in this survey, we differentiate threats aimed at data reconstruction from PIAs and classify them as Inversion Attacks. 

Attackers in the study \cite{melis2019exploiting} use auxiliary data and trained classifiers to determine if observed model updates are generated from a dataset that includes the target property. Considering fully connected neural networks (FCNNs) are invariant under the permutation of nodes when represented using matrices, study \cite{ganju2018property} introduces PIAs in white-box settings and shows good inference performance in shallow-FCNNs. The study \cite{274683} further extends PIAs to logistic regression (LR), long short-term memory networks (LSTM), and graph convolutional networks (GCN) across multiple domains. It demonstrates that even in black-box settings, PIAs can still achieve accuracies exceeding 60\%.

\subsubsection{Representative Inference Attacks (RIA)}
Representative Inference Attacks indicate the malicious generate dummy outcome that is representative of the victim's training data \cite{hitaj2017deep, wang2019beyond}. The study \cite{hitaj2017deep} first introduces record-level RIAs in collaborative learning scenarios and demonstrates that GAN-based attacks can effectively compromise against Distributed Selective SGD (DSSGD \cite{shokri2015privacy}). However, subsequent research \cite{wang2019beyond} reveals its performance degradation in FL settings due to the averaging of updates and introduces the mGAN-AI framework. This framework enhances GANs with a multitask discriminator capable of simultaneously assessing the category, reality, and client identity of input samples. Compared to \cite{hitaj2017deep}, mGAN-AI enables the generator to recover representative private data for specific users.

\subsection{Inversion Attacks}
Although FL allows clients to only upload gradients during model training, it has been found that some properties of the training data are implicitly carried by these gradients, which could be revealed through inverse attacks. DLG \cite{zhu2019deep} first introduces gradient inversion in FL computer vision areas. Specifically, the DLG generates random dummy input samples $x'$ and labels $y'$ and corresponding dummy gradients $G'$. $x'$ and $y'$ are Iteratively optimized through L-BFGS to minimize the difference between dummy and benign gradient. Formally, the objective function of DLG shown as follows:

\begin{equation}
    \mathop{\arg\min}\limits_{x',y'} ||G'-G||^2
\end{equation}

Considering the high-dimensional direction of gradients carried important information, study \cite{geiping2020inverting} introduces cosine similarity in the loss function to find images that lead to a similar change in model prediction as the ground truth instead of finding images with a gradient that best fits the observed gradient. The objective function is shown as follows, where \textit{TV} represents the Total Variance regularization.

\begin{equation}
    \mathop{\arg\min}\limits_{x'}(1-\frac{G'\cdot G}{||G'||\cdot ||G||}+\alpha \cdot TV(x'))
\end{equation}

Although DLG \cite{zhu2019deep}, IG\cite{geiping2020inverting} show effectiveness on single image and small batch size training scenarios, their inversion performance degrades when facing large batch size ($>8$). To further improve the attack effectiveness, GGL \cite{Li_2022_CVPR} leverages a generative model trained on public datasets as the learned natural image prior to ensure the reconstructed image quality. Additionally, GradInversion (GI) \cite{yin2021see} integrates Fidelity and Group Consistency regularization into the objective function and effectively increases the inversion batch size to 48. These are designed to ensure that the generated images closely resemble real images and to penalize any dummy images that deviate from a “consensus” image, respectively. GradViT \cite{hatamizadeh2022gradvit} extends the GI approach to Vision Transformer (ViT) models, demonstrating that ViTs are significantly more vulnerable to inversion attacks than the previously studied CNNs due to the attention mechanism. 

We collect the images from study \cite{review2024, yin2021see} and compare the ImageNet batch gradient inversion outcomes under different inversion attacks in Figure \ref{compare2}.

\subsection{Eavesdropping}
Network eavesdropping involves an attacker intercepting and capturing data packets transmitted over a network using their own devices, with the intent to analyze the contents of these packets to get private information \cite{zou2013eavesdropping}. Since model training under the FL scheme requires regular communication between the server and clients, FL is naturally vulnerable to eavesdropping attacks. This vulnerability is particularly pronounced when the communication between participants and the server is in plaintext or employs a weak encryption method \cite{mothukuri2021survey,wang2023energy}. Since the messages transmitted are usually models or gradients, the attacker may further leverage inference or inversion attacks to decipher the eavesdropped messages to understand the private information \cite{ma2020safeguarding,9839650,xu2021else}. Study \cite{10051719} considers the eavesdropping attacks in unmanned aerial vehicles (UAVs) scenarios, demonstrating that eavesdroppers can deduce the raw data from the shared parameters and compromise participants' privacy.

\section{Defence Against Utility-Harmed Attacks}\label{SEC5}
As the diversity and complexity of adversarial attacks against FL increase, new defenses are being developed to counteract their malicious impacts. These defense frameworks generally follow the setting that the server is honest and can play the defender role. As some defense methods are effective against multiple types of attack categories, we survey these FL defense frameworks from different technologies instead of grouping them according to the attack category defended. Table \ref{defense} provides the overview of these methods.

\newcolumntype{C}[1]{>{\centering\arraybackslash}m{#1}}
\begin{table*}[htbp]
\caption{Illustration and comparison of different defense methods for utility-harmed attacks.\\}
\begin{center}
\begin{tabular}{C{2.4cm}|C{1cm}|C{1cm}|c|C{3cm}|C{5.5cm}}
\toprule[1pt]
\midrule
\textbf{Defense Methods}&\textbf{Trusted Server}&\textbf{Defender}&\textbf{Preformed Phase}&\textbf{Technique}&\textbf{Remark}\\
\midrule
Data Sanitization &$\checkmark$&Client, Server&Before training&Remove suspected data&Suffering from privacy issue, applicable to attacks related to poisoned data.\\
\midrule
Anomaly Detection &$\checkmark$&Server&Before aggregation&Filter suspected identity&Suffering from privacy and non-IID issues.\\
\midrule
Adversarial Training &$\checkmark$&Client&During training&Robustness enhancement&Could degrade the model performance.\\
\midrule
Model Compress&$\checkmark$&N/A&N/A&Attack surface reduction&Also mitigates the communication bottleneck but may degrade the model performance.\\
\midrule
Byzantine-Resilient Aggregation&$\checkmark$&Server&During aggregation&Mitigate impacts carried by gradient outliers&Performance could be limited by assumptions and knowledge of adversary.\\
\midrule
Blockchain FL&$\times$&Client&N/A&Fully decentralized mechanism&Can address the single point of failure issue but cannot inherently defend against attacks.\\
\midrule
\bottomrule[1pt]
\end{tabular}
\label{defense}
\end{center}
\end{table*}

\subsection{Data Sanitization}
Data sanitization \cite{4531146, li2023martfl, 9248056} has been introduced to remove malicious and suspicious data before the training process and consequently mitigate crafted-data-related threats, including data poisoning and evasion attacks. The research \cite{4531146} introduces a novel training pipeline that incorporates a data sanitization phase. During each iteration, this phase creates multiple models, termed micro-models, based on subsets of the training data. These micro-models are used to produce provisional labels for each training input and consequently combined in a voting scheme to determine which parts of the training data may be potentially attacked. To filter out noise-labeled data, FedDiv \cite{li2024feddiv} maintains a global filter. This filter is trained, exchanged, and aggregated similarly to the global model and supports local data filtration. From the data shuffling perspective, study \cite{9248056} proposes a byzantine-resilient matrix-vector (MV) multiplication and further proposes an algorithm based on the data encoding process and error correction over real numbers to combat various adversarial attacks.

Although data sanitization can filter out some potential attacked data and keep training data clean, it relies on directly accessing the local data of clients, which may raise privacy issues. A recent work \cite{koh2022stronger} indicates that data sanitization can only work in
cross-silo FL with strong authentication and be performed locally by the client itself. However, this framework \cite{koh2022stronger} assumes the trustworthiness of the clients, which may not be applicable to the adversarial FL environment. 

\subsection{Anomaly Detection}
To perform threats, attackers unconsciously exhibit different patterns from benign clients, including abnormal behaviors or data patterns. To this end, anomaly detection has been proposed to identify suspicious through statistical or behavior analysis technology \cite{li2019abnormal, meng2021vadaf,  hong2023federated}. We follow the study \cite{Gao2022survey} and divide the anomaly detection of FL in client and data from two perspectives. 

\textit{A) Client Anomaly Detection}\\
Under FL, Client Anomaly Detection could be performed based on the information from the communication  (IP address, User ID, Time Stamp \cite{li2023enhancing}) or application layer, namely client model updates. By employing a pre-trained auto-encoder model, study \cite{li2019abnormal} introduces Auto-encoder-Based Anomaly Detection to detect abnormal model weight updates from the clients and consequently erase the negative impact carried by the potential attackers. Inspired by spectral anomaly detection that distinguishes abnormal data instances by capturing the normal data features, research \cite{li2020learning} designs a novel robust FL framework. The spectral anomaly detection model is trained through a centralized training process and applied in the following learning iteration. Since the detection threshold is adjusted dynamically in each communication round, it becomes challenging for the adversary to understand the framework, ensuring they can be effectively excluded from the aggregation. Recent research \cite{meng2021vadaf} presents a visualization scheme for anomaly detection, introducing VADAF. Specifically, VADAF captures the intricate dynamics of the FL training process, aiding in investigating various client issues and assessing the severity of anomalous clients.

\textit{B) Data Anomaly Detection}\\
Anomalous data, potentially introduced during the data sampling process, can adversely affect the model's training performance. Consequently, data anomaly detection has been introduced to identify outliers in the dataset or values that are far from the normal data feature values. Research \cite{chen2018automated} proposes a two-phase iterative adversarial detection method to identify software samples in the malicious application detection system that have been subjected to poisoning attacks. Research \cite{kieu2019outlier} proposes an anomaly detection method for time series datasets based on recursive auto-encoding, which reduces the impact of overfitting to anomalies. To further improve robust and efficient anomaly detection in unsupervised time series scenarios, a variational recurrent encoder model \cite{kieu2022anomaly} can separate anomalies from normal data without relying on anomaly labels. However, this operation requires direct access to clients' data or transferring trust to participants, leading to limitations similar to those in Data Sanitization.

\subsection{Adversarial Training}
During the model training process, a minor and well-designed perturbation can be manually introduced to enhance the model's robustness, namely Adversarial Training. 

Research \cite{tramer2018ensemble} first introduces ensemble adversarial training in centralized learning settings, which augments training data through transferring perturbations from other pre-trained models. Studies \cite{shah2021adversarial, hong2023federated, chen2021certifiably} further extend the adversarial training in decentralized settings. Specifically, study \cite{shah2021adversarial} applies adversarial training in the FL environment to reduce model drift and accelerate model convergence. Nevertheless, as the perturbation introduction inevitably affects the classification accuracy, the performance carried by adversarial training may not be stable against complex black-box attacks. On the other hand, adversarial training relies on a massive volume of training datasets, which increases the cost of computation resources on the local devices. Especially in cross-device FL environments with multi-participants, lightweight clients can not afford the high costs of adversarial training. Research \cite{hong2023federated} proposes a novel learning scheme that propagates adversarial robustness from high-resource users, who can undertake adversarial learning, to low-resource users throughout the FL process. Research \cite{chen2021certifiably} investigates the effectiveness of defencing Evasion Attacks through adversarial learning. Specifically, Gaussian noise is used to smooth the training data by including adversarial data in the training dataset.

\subsection{Model Compression}
The large model size enables the DL model to achieve a good learning performance but also introduces both communication challenges and a broadened attack surface in the FL environment. To reduce communication overhead and parameter redundancy, model compression technology has been widely investigated by recent studies \cite{phuong2019towards, li2019fedmd, lin2020ensemble}.

As one of the model compression technologies, knowledge distillation has been incorporated into transfer FL in research \cite{li2019fedmd}, which introduces a general framework named FedMD. FedMD allows clients with different computational resources to design different network structures, protecting data privacy and enhancing the performance of local models. While FedMD improves the local model performance, the participants are required to devote a portion of data privacy to form a shared dataset. Following FedMD \cite{li2019fedmd}, research \cite{lin2020ensemble} leverages unlabeled data from local model outputs to achieve model fusion, which further accelerates model convergence and mitigates data privacy leakage.

\begin{table*}[htbp]
\caption{Illustration of the techniques, requirements, and limitations of various byzantine-robust aggregation methods.\\}
\begin{center}
\begin{tabular}{c|c|c|c}
\toprule[1pt]
\midrule
\textbf{Aggregation}&\textbf{Algorithm}&\textbf{Technique}&\textbf{Requirements \& Limitations}\\
\midrule
& (Multi) Krum\cite{blanchard2017machine}&Euclidean distance&Knowledge of adversary amount assumption, non-IID issue\\
&Trimmed-Mean \cite{yin2018byzantine}&Mean value&Knowledge of adversary amount assumption, non-IID issue\\
&Median \cite{yin2018byzantine}&Median value&Non-IID issue\\
&Bulyan \cite{guerraoui2018hidden}&Euclidean distance+Mean value&Knowledge of adversary amount assumption, non-IID issue\\
Statistical Feature&RFA \cite{pillutla2022robust}&Geometric median&Applicable to a few poisoning attacks\\
-based&Sniper \cite{cao2019understanding}&Maximum clique problem&Applicable to attacks bring large perturbation\\
&AFA \cite{munoz2019byzantine}&Cosine similarity+Adaptive mean&Benign clients' gradients should be close, non-IID issue\\
&LICM-SGD \cite{yang2019byzantine}&Lipschitz constant&Non-IID issue\\
&FoolsGold \cite{fung2018mitigating}&Cosine similarity&Applicable to sybil attacks\\
&MeaMed \cite{xie2018generalized}&Mean+Median&Non-IID issue, Curse of dimensionality\\
&Byrd-SAGA \cite{wu2020federated}&SAGA+Geometric median&Non-IID issue\\
\midrule
& AVBA \cite{wang2020model}& Client model accuracy& Applicable to model poisoning attacks\\
& FL Trust \cite{cao2020fltrust}& Cosine similarity+Clipping& Clean\&IID evaluation dataset, vulnerable to specific attacks\\
 &Sageflow \cite{park2021sageflow}& Loss value& Clean\&IID evaluation dataset, vulnerable to specific attacks\\
&Contribution-wise \cite{LI2024120475}& Cosine sim/Loss value+Clipping& Clean\&IID evaluation dataset\\
Performance-based&Zeno \cite{xie2019zeno}& Loss value (decrease)& A sampling method for collecting evaluation dataset\\
&DGDA \cite{cao2019distributed}& Loss value variant & Applicable to attacks bring large perturbation\\
&SCL \cite{zhao2020shielding}&Client cross validation& Extra communication and computation cost\\
&PDGAN \cite{zhao2020pdgan}&GAN generates evaluation data& Extra training iterations\\
&HSCS FL \cite{li2023honest}&Client model accuracy on class& Applicable to label flipping attacks\\
\midrule
&Byzantine ClusterFL \cite{sattler2020byzantine}&Cosine similarity& Malicious clients should be grouped as the largest group\\
&Mini FL \cite{li2023enhancing}&Non-IID feature+K means& Non-IID feature should be reliable and explicit\\
&RFL in HE \cite{ghosh2019robust}&Trimmed-K means& Applicable to attacks bring large perturbation\\
Clustering-based&Robust-DPFL \cite{qi2024towards}&Mean value+K means& Mainly applicable to Attack-DPFL \cite{qi2024towards} attacks\\
&Fair attack detection \cite{singh2023fair}&Feature+Gaussian mixture model&Clients have to sacrifice a part of privacy\\
&ZeKoC \cite{chen2020zero}&Euclidean distance+K means&Evaluation dataset, Applicable to model poisoning attacks\\
&DAFLA \cite{tolpegin2020data}&Principal component analysis&Non-IID issue, Applicable to label flipping attacks\\
\midrule
\multirow{2}{*}{Training function}&RSA \cite{li2019rsa}&Regularization-based loss function& Assumptions of general stochastic gradient-based methods\\
\multirow{2}{*}{optimization-Based}&HLMA \cite{zhao2024huber}&Huber loss variant& (extreme) Non-IID issue\\
&RLR \cite{ozdayi2021defending}&Learning rate adjustment&Applicable to backdoor attacks\\
\midrule
\bottomrule[1pt]
\end{tabular}
\label{tab4}
\end{center}
\end{table*}

Pruning is another model compression technique that can eliminate redundant or poisoned model neurons. Research \cite{liu2018fine} introduces a new pruning technique through trimming and fine-tuning the model returned by attackers to defend against backdoor poisoning attacks on the training set and communication bottleneck attacks. Recent research \cite{jiang2022model} proposes a PruneFL method with adaptive and distributed parameter pruning in FL settings. Through employing PruneFL, the model can achieve a similar learning performance while reducing computational costs and shortening the learning time. Inspired by the coding theory \cite{van1998introduction}, \cite{chen2018draco} proposes the DRACO framework for robust distributed training. In the DRACO \cite{chen2018draco} framework, each client evaluates redundant gradients, which the parameter server subsequently employs to counteract the impact of adversarial updates. Similarly, DETOX \cite{rajput2019detox} has also been introduced based on redundancy to provide a more robust gradient aggregation. DETOX algorithm \cite{rajput2019detox} includes two steps: a filtering step to reduce the effect of byzantine nodes by using limited redundancy and a hierarchical aggregation step that could be used in tandem with other existing robust aggregation methods. CrowdGuard \cite{riegercrowdguard} further enhances the pruning scheme by introducing client cross-validation. Specifically, suspicious client models are marked as malicious and pruned during aggregation to prevent backdoor attacks.

\subsection{Byzantine-Resilient Aggregation}\label{Byzan R}
Linear combination rules, including averaging, have been demonstrated to lack resilience against byzantine attacks \cite{blanchard2017machine}. In particular, a single malicious worker can corrupt the global model and even prevent global model convergence. To defend against various attacks, byzantine-resilient aggregation methods have been proposed to replace the averaging aggregation and enhance FL robustness. 

\subsubsection{Statistical Feature-Based Robust Aggregation}
When introducing perturbation in the model updates, malicious updates are unavoidable to deviate more from other normal updates in FL. Based on this insight, some research designs statistical-based aggregators on the server side to aggregate the most honest clients' gradients and keep the global model away from potential malicious updates. For instance, Krum (multi-Krum) \cite{blanchard2017machine} regards the client (could be a small number of clients) that is closest to all participants as the most honest and uses its (them) gradient as the aggregation outcome. Specifically, Krum first calculates the L2 (Euclidean) distance between the gradient to others and consequently ranks the distance. The gradient achieving the shortest distance value will act as the global gradient and update the global model. Trimmed mean \cite{yin2018byzantine} and Median \cite{yin2018byzantine} enhance the FL robustness by selecting the mean and median value of each gradient parameter; these coordinates are subsequently organized as the aggregation outcome. To avoid the aggregation outcome being impacted by the extreme value, Trimmed mean first discards a preset percentage of extremum. Bulyan \cite{guerraoui2018hidden} combines the Krum and Trimmed-mean algorithms. Bulyan first selects a number of gradients as candidates by repeatedly performing Krum. and removes them from the candidate pool. Then, Bulyan performs Trimmed-mean among the candidates and generates the outcome to update the global model.

However, these previously mentioned methods operate under the strong assumption that the defender is aware of the adversary's volume and distribution, an assumption that may be challenging to apply in practical scenarios. Research \cite{cao2019understanding} addresses this limitation by proposing the Sniper scheme. Sniper \cite{cao2019understanding} does not require prior knowledge of the number of malicious users and identifies benign local model updates by solving the Maximum Clique Problem (MCP). Similarly, study \cite{moreno2012unifying} relies on the Markov model with median and cosine similarly to discard the suspicious local model updates in each iteration and does not require the knowledge of the adversary either. 

\subsubsection{Local Model Performance-Based Robust Aggregation}
To evaluate the honesty of clients' model updates and limit the participation of suspicion, some recent studies introduce an evaluation dataset to decide the aggregation weight of each client. For instance, FL Trust \cite{cao2020fltrust} proposes a novel solution by keeping a server-version model. Under FL Trust, the server first collects a small and clean dataset as the training dataset. Then, the server normally keeps and trains the model (called the server-version model) in each iteration. Once the clients' gradients are received, the server calculates the cosine similarity between them and the server-version gradient; the gradient achieving high cosine similarity will subsequently gain a high weight in aggregation. On the other hand, FL Trust normalizes the magnitudes of clients' gradients through the server-version model update to defend against scaling attacks. Similarly, Sageflow \cite{park2021sageflow} evaluates the entropy of each gradient received through the evaluation dataset and subsequently divides the weight to different clients based on the entropy generated. Sageflow also considers the volume of training data in each client; the client owning a large data size can gain a large weight in the aggregation and vice versa. Research \cite{xie2019zeno} proposes a score-based ranking mechanism, namely Zeno. Zeno suspects clients that are potentially defective, which is tolerant to an arbitrary number of attackers with at least one benign client. Based on Zeno, Zeno++ \cite{xie2020zeno++} has been introduced to further mitigate the communication bottleneck, allowing for asynchronous communication. Research \cite{cao2019distributed} generates the gradient of noise based on a small clean dataset. The noise gradient is subsequently compared with the clients' model updates to identify and filter out suspicious gradients. 

Although the existing byzantine-robust aggregation can achieve robustness under particular assumptions like independent and identically distributed (IID) distribution of overall training data, the knowledge of clients' data distribution, or a set of clean public data, these assumptions can not be guaranteed in the practical FL scenarios, which usually results in degrading defense performance and leaves opportunities for state-of-the-art attacks to poison the model. 

\subsubsection{Clustering-Based Robust Aggregation}
In the FL context, clients may have different preferences, making it challenging for a single model to fit all clients' data-generating distributions at the same time. Based on this insight, Clustered Federated Learning (ClusterFL) \cite{sattler2020clustered} has been introduced, leveraging the geometric properties of the FL loss surface to group the client population into clusters with jointly trainable data distributions. The study \cite{sattler2020byzantine} further extends ClusterFL in the adversarial environment. Under standard ClusterFL \cite{sattler2020clustered}, cosine similarity is used to determine the cluster of each client, the client distribution ensures that the cosine similarity is maximized between clients within the same cluster and minimized between different clusters. Byzantine-robust ClusterFL \cite{sattler2020byzantine} employs the same method to divide clients but considers the largest cluster to be benign while treating all other clusters as malicious. As a result, the smaller clusters are discarded during aggregation. 
The study \cite{li2023enhancing} considers the non-independent and identically distributed (non-IID) nature of FL as the primary factor contributing to a large attack surface, and proposes Mini-FL to achieve attack surface reduction. Specifically, Mini-FL identifies time, geography, and user features as the main sources of non-IID data and sets the grouping rules according to the characteristics of these features. In each iteration, the received gradients are assigned to different groups and robustly aggregated, respectively. The aggregation outcomes from each group are then further combined into a global gradient, which supports the model update. Besides non-IID, study \cite{qi2024towards} finds that the Differential Privacy (DP) introduced after local training also expands the overall attack surface and enables adversarial to perform attacks. 
To this end, Robust-DPFL \cite{qi2024towards} has been introduced to distinguish poisoned gradients from clean ones. Since malicious gradient elements are naturally larger than benign ones, Robust-DPFL first averages the element values and then uses k-means clustering to divide the received gradients into clean (with smaller values) and poisoned (with larger values) categories.

\subsubsection{Training Function Optimization-Based Robust Aggregation}
Different from the aforementioned aggregation rules that filter or limit the suspicious based on their gradient, training function optimization-based robust aggregation leverages the DL loss function optimization to achieve byzantine resilience. Compared with other robust aggregations, training function optimization-based robust aggregation is still in the early research stage. Research \cite{li2019rsa} introduces regularization in the loss function to prevent the local model from deviating excessively from the global model during training. Instead of relying on a fixed clipping norm, study \cite{andrew2021differentially} proposes a method with dynamic clip value at the specific quantile estimated online. Recent work \cite{zhao2024huber} generalizes the Huber loss definition and leverages the optimal Huber loss to organize the loss function. Although this approach provides a theoretical guarantee under IID and moderately heterogeneous client environments, the guarantee may not hold in extreme non-IID settings.

\subsection{Other Emerging Defense Frameworks}
Although most byzantine-resilient FL methods \cite{kairouz2021advances} follow the setting that the server (service provider) acts as the defender role, some works consider defending the poisoning attacks from the client's perspective. A recent work \cite{sun2021fl} proposes a novel algorithm to enhance FL robustness from the participants' side, namely White Blood Cell for Federated Learning (FL-WBS). This study \cite{sun2021fl} first introduces the Attack Effect on Parameter (AEP) to evaluate the model poisoning attacks' impact on global model parameters and subsequently introduces the FL-WBS framework to prevent AEP from being hidden in the kernels of Hessian matrices on benign devices. SVeriFL\cite{gao2023sverifl} leverages BLS signature and homomorphic encryption technology to verify the integrity of the parameters uploaded by the participants and the correctness of the results returned by the server.

Owing the anonymity and distributed characteristics, enhancing FL robustness through blockchain has become an appealing research trend \cite{nguyen2021federated, qu2022blockchain}. Research \cite{kim2019blockchained} considers mitigating the one-point failure (server be controlled by malicious) of FL by introducing blockchain technology in the current FL algorithm. Specifically, research \cite{kim2019blockchained} proposes BlockFL, which replaces the server with a blockchain to relieve the server security issue and achieve decentralization. In each iteration, the clients normally train the model and then broadcast the gradients to the miners. The miners cross-verify the gradients and generate the candidate block until a preset block size is achieved. Once the block is published, the clients download the block and locally compute the global model update and consequently utilize the outcome to update the model. Study \cite{9399813} proposes a reputation-based FLchain system, namely BAFL, to achieve asynchronous FL strategy and maintain FL robustness. Under BAFL \cite{9399813}, the training time, training data size, gradients correlation, and global update cheating times are generated by the entropy weight method and are recorded in the blockchain to achieve long-term trust.

Some work \cite{liang2018detecting, zhao2020pdgan, ross2018improving, xie2023unraveling} consider introducing the defense framework from traditional centralized ML scenarios to FL settings. For instance, PDGAN \cite{zhao2020pdgan} generates testing datasets through GAN technology to defend against Data Poisoning Attacks. Research \cite{zhuang2014towards} proposes Moving Target Defense to increase attacking cost and complexity by changing a system to reduce or move the attack surface available for exploitation by the adversary. However, most of these studies \cite{zhuang2014towards, ross2018improving} may show effectiveness in Horizontal FL (HFL) settings but face new challenges when deploying in Vertical FL (VFL) scenarios.

\section{Defense Against Privacy-harmed Attacks}\label{SEC6}
Unlike utility-oriented defense frameworks that generally assume a trusted central server, defenses against privacy-oriented attacks can be implemented by either the server or the clients, depending on different settings. Today, various defense frameworks have been proposed to mitigate privacy concerns, ranging from cryptography-based methods to fully decentralized architectures. In this section, we survey these defense solutions, introduce their techniques, and compare their advantages and limitations.
\subsection{Differential Privacy}
Differential Privacy (DP) has been proposed as a privacy protection data-sharing method. By introducing artificial noise, DP guarantees statistical queries on a database without disclosing information specific to individuals \cite{dwork2014algorithmic}. In the FL system, DP provides a privacy guarantee that clients' data and information cannot be recovered during the training and user access phases.

\subsubsection{Centralized Differential Privacy (CDP)}
When the server is trusted, noise can be added centrally to the aggregated local gradients (known as CDP) to ensure record-level privacy for the entire dataset of all participants. Formally, the definition of CDP is presented as follows:

\textbf{\textit{Definition 1 (($\epsilon$,$	\delta$)-CDP \cite{dwork2014algorithmic}):}} The algorithm $\mathcal{A}$ satisfies ($\epsilon$,$\delta$)-DP, if any pair of adjacent datasets $D$, $D'\in \mathcal{D}^n$ and outputs $S \in $range$(\mathcal{A})$ we have:
\begin{equation}
    Pr[\mathcal{A}(D)\in S]\leq e^\epsilon \cdot Pr[\mathcal{A}(D')\in S]+\delta\nonumber
\end{equation}

Here, $\epsilon> 0$ controls the level of privacy guarantee in the worst case, and the factor $\delta \geq 0$ represents the failure probability that the privacy guarantee does not hold. In practice, the value of $\delta$ should be negligible, such that $\delta \ll 1/||D||$.

For the first time, the study \cite{geyer2017differentially} introduces CDP in FL to conceal clients’ contributions during training. Subsequently, DP-FedAvg and DP-FedSGD \cite{mcmahan2018learning} extend CDP-FL to LSTM language models, achieving results that are quantitatively and qualitatively similar to those of un-noised learning algorithms. To reduce the impact of the DP mechanism, DP-FedSAM \cite{foret2020sharpness} leverages Sharpness-Aware Minimization (SAM) optimization to search for areas of low loss in the loss landscape and improve the model's generalization ability. By sparsifying local updates and excluding the softmax layer from uploading, studies \cite{cheng2022differentially} and \cite{xu2023learning} also achieve a reduction in the volume of injected noise and enhance the performance of CDP-FL. Considering that both data features and label spaces could be regarded as private in VFL scenarios, HDP-VFL \cite{wang2020hybrid} introduces Gaussian noise on both the server and client sides to ensure the data confidentiality of VFL participants.

While CDP-FL demonstrates effectiveness in privacy protection, it faces limitations from two main perspectives \cite{9945997}: (i) CDP-FL algorithms generally rely on a large number of participants to achieve convergence, learning scenarios with a small number of participants may result in significant performance degradation. (ii) The assumption of a trustworthy server may not hold in many real-world settings.

\subsubsection{Local Differential Privacy (LDP)} Different from CDP, LDP introduces the DP noise from the participant side to guarantee data privacy. As the gradient left from the participant surface is noised, LDP does not assume a trusted server and potential defenses against eavesdropping.  Formally, the definition of LDP is presented as follows:

\textbf{\textit{Definition 2 (($\epsilon$,$	\delta$)-LDP \cite{duchi2013local}):}} The algorithm $\mathcal{A}$ satisfies ($\epsilon$,$\delta$)-DP, if any input $d$, $d$ and $\forall t\in$range$(\mathcal{A})$ we have:

\begin{equation}
    Pr[\mathcal{A}(d)=t]\leq e^\epsilon \cdot Pr[\mathcal{A}(d')=t]+\delta\nonumber
\end{equation}
Early LDP-FL works \cite{shokri2015privacy, truex2020ldp} required clients to locally introduce noise to model parameters before releasing them to the server. However, with the explosive growth of model size and dimensions, the privacy budget allocated to each dimension decreases, preventing these methods from providing an expected privacy guarantee. On one hand, some studies propose new mean estimation techniques for high-dimensional data to enhance the performance of LDP-FL, including Piecewise Mechanism (PM) \cite{wang2020hybrid, zhao2020local} and Hybrid Mechanism (HM) \cite{wang2020hybrid}. On the other hand, studies that leverage dimension sampling have also shown effectiveness in mitigating the curse of dimensionality in LDP-FL \cite{fu2024differentially}. For instance, PPeFL \cite{wang2023ppefl} introduces a filtering and screening method through an exponential mechanism (EM). Based on the contribution of weight parameters to the neural network, PPeFL filters out the better parameters using EM, thereby addressing the parameter dimension dependence problem to some extent when LDP is applied to complex ML tasks.
\\

Because of the different perturbation introduced strategies, CDP and LPD show different robustness when facing the same threats. Study \cite{naseri2020local} compares these two methods and finds that: (i) Both LDP and CDP protect against white-box MIA and do not necessarily incur a high cost on utility. (ii) LDP does not work against PIA. Although CDP can defend against such attacks, it suffers from a significant model utility degradation. We follow the study and compare the representative DL-FLs in Table \ref{tab4}. Meanwhile, study \cite{fu2024differentially} provides a detailed table of DP-FLs from other perspectives.

\begin{table*}[htbp]
\caption{Illustration and comparison of different DP-FL methods.}
\begin{center}
\begin{tabular}{c|c|c|c|c|c|c}
\toprule[1pt]
\midrule
\textbf{DP Category}&\textbf{Algorithm}&\textbf{FL settings}&\textbf{Non-IID}&\textbf{Perturbation}&\textbf{Model}&\textbf{Learning Task Dataset}\\
\midrule
\multirow{5}{*}{CDP}&DP-FL\cite{huang2020dp}& HFL &\halfcirc& Gaussian & CNN & MNIST, Cifar-10, Iris\\
&UDP-FL\cite{wei2021user}& HFL &\halfcirc& Gaussian &SVM, MLP, CNN &IPUMS-US, MNIST, Cifar-10\\
&Federated f-DP\cite{zheng2021federated}& HFL&\emptycirc& Gaussian &CNN & MNIST\\
&FEDMD-NFDP\cite{sunfederated}& TFL&\emptycirc&Sampling &MLP & MNIST, Cifar-10, -100 (variants)\\
&DynamicPFL\cite{yang2023dynamic} &HFL &\emptycirc &Gaussian &CNN &Cifar-10, FEMNIST, SVHN\\
\midrule
&2DF-IDS\cite{zheng2021federated}& HFL (DFL) &\emptycirc &Gaussian &DNN & Edge-IIoTset\\
&VFL-Chain\cite{smahi2024vfl}& VFL (DFL) &\fullcirc &Laplace &N/A &MNIST, BelgiumTS\\
LDP&DP-FedSAM\cite{shi2023make}& HFL &\emptycirc &Gaussian &CNN, ResNet-18 & EMNIST, Cifar-10, -100\\
&FL2DP\cite{gu2023fl2dp}& HFL &\fullcirc &Exponential &CNN & MNIST, Cifar-10\\
&ProxyFL\cite{kalra2023decentralized}& HFL (DFL)&\halfcirc &Gaussian &LeNet5, MLP, CNN &Fashion-, MNIST, Cifar-10\\
\midrule
\multirow{2}{*}{Hybrid}&iDP-FL\cite{zhang2024idp}& HFL &\emptycirc &Gaussian &MLP & Fashion-, MNIST, Cifar-10\\
&PPFL\cite{mo2021ppfl}& HFL &\fullcirc &Gaussian, Laplace&MLP &MNIST\\
\midrule
\multicolumn{7}{c}{{\fullcirc}\ Not applicable, \ \  \halfcirc \ Applicable to limited settings, \ \  \emptycirc \ Applicable}\\
\bottomrule[1pt]
\end{tabular}
\label{tab4}
\end{center}
\end{table*}

\subsection{Decentralized Learning Scheme}
\subsubsection{Blockchain}
Since curious central servers are one of the main sources of privacy-oriented attacks, recent studies have considered leveraging blockchain as a decentralized alternative to replace the central server \cite{sameera2024privacy}. 

\begin{figure} [h]
    \centering
    \includegraphics[width=1.0\linewidth]{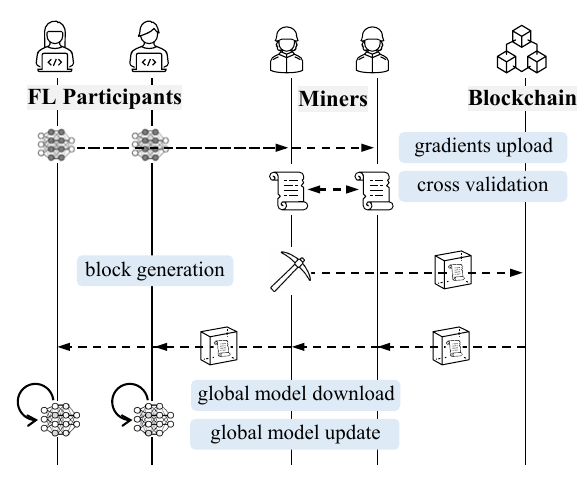}
    \caption{The learning process of blockchain-based FL.}
    \label{blockfl}
\end{figure}

The blockchain-based FL (BCFL) \cite{kim2019blockchained, singh2022framework, kumar2021blockchain, li2020blockchain} generally follow the learning process shown as the Figure \ref{blockfl}. Specifically, the clients locally train the models and upload gradients to associated miners. Miner broadcasts received gradients and records in the candidate block after cross-verification. Once a preset condition is met, each miner runs a cryptographic proof, and the first successor propagates its block. Finally, all participants download the block from the associated miners and update the global model accordingly. As the whole learning process does not require centralized collected gradients, blockchain-based FLs are potentially robust against inversion attacks.

To enhance scalability, studies  \cite{guduri2023blockchain, ramanan2020baffle, zhang2023privacy} introduce smart contracts, which serve as executable code embodying mutual agreements between two or more parties in blockchain-based FL systems. For instance, the study \cite{guduri2023blockchain} utilizes active smart contracts to automatically monitor, organize, and facilitate secure data transfer, validating the system's efficacy. Similarly, BAFFLE \cite{ramanan2020baffle} uses smart contracts to decompose global parameters into different blocks, employing strategies to perform model aggregation and update tasks. To mitigate the storage bottleneck of blockchain, PriModChain \cite{zhang2023privacy} introduces the Inter Planetary File System (IPFS) to blockchain-based FL. Under PriModChain, IPFS provides extensive decentralized storage, while the blockchain stores only the target addresses of the IPFS.

Besides enhancing privacy, blockchain also benefits the FL system from an incentive perspective. The study \cite{qu2020blockchained} proposes a completely decentralized framework using blockchain technology to reward members with high-performance or high-quality data, encouraging voluntary contributions and potentially enhancing model performance.

\subsubsection{Decentralized Federated Learning (DFL)}
To mitigate the privacy leakage risk of FL, another research pipeline is that transfer the centralized FL to decentralized (shown as Figure \ref{CDFL}). The study  \cite{lalitha2018fully} first proposes a DFL scheme, where clients directly update their beliefs by aggregating information from their one-hop neighbors to learn a model that best fits the observations. Study \cite{kalra2023decentralized} further introduces the KD in DFL and proposes a ProxyFL scheme. Under ProxyFL, each client owns both a student model and a teacher model, which are responsible for learning knowledge and model exchange, respectively. Since the student model is not shared during the training process, ProxyFL provides property (model as property) and privacy protection for the model architecture and trained parameters.

While DFL offers a novel solution for addressing FL privacy concerns, existing algorithms generally assume that clients have reliable and stable networks and cannot tolerate stragglers, which hinders their deployment in the real world.
\subsection{Homomorphic Encryption}
Homomorphic encryption (HE) is a form of encryption that enables operations to be performed on encrypted data, and guarantees the encrypted computations results identical to the output of operations performed on the unencrypted data. Formally, HE satisfies the following equation \cite{Gu2023Survey}:

\begin{equation}
    Dec[Enc(g_1)\odot Enc(g_2)] = g_1\odot g_2
\end{equation}
Here, $Enc(\cdot)$ and $Dec[\cdot]$ represent encryption and decryption, respectively, and $g$ represents the plaintext, which is typically the gradient in the FL context. Based on the limitations of the operations $\odot$ (addition $\oplus$ or multiplication $\otimes$), HE could be categorized into: Fully HE (FHE), Somewhat HE (SWHE), and Partially HE (PHE). We compare the three different types of HE and summarize their characteristics in Table \ref{tab5}.

\newcolumntype{C}[1]{>{\centering\arraybackslash}m{#1}}

\begin{table*} [h]
    \centering\caption{Illustration of homomorphic encryption methods and applied HE-FL schemes.}
\begin{tabular}{c|c|c|c|c|c}
\toprule[1pt]
\midrule
\textbf{Homomorphic Encryption}&\textbf{Operation times}&\textbf{$\oplus$}&\textbf{$\otimes$}&\textbf{Computation cost}&\textbf{Summary}\\
\midrule
FHE \cite{li2024secure, kumar2024revamping, sav2021poseidon} &Unlimited& \checkmark &\checkmark& High& \makecell[c]{High privacy protection, but the high computation cost\\ limits it be applied in the real world applications.}\\
\midrule
SWHE \cite{zhou20222d ,zhang2020batchcrypt}&Limited& \checkmark &\checkmark& Medium&Operations may introduce noise, applicable to specific scenarios.\\
\midrule
PHE (AHE) \cite{phong2017privacy, hao2019towards, chai2020secure}&Unlimited& \checkmark & $\times$ &\multirow{2}{*}{Medium to Low}&\multirow{2}{*}{\makecell[c]{Simple in design but limited in functionality,\\ applicable across various scenarios.}} \\
PHE (MHE) \cite{fang2021privacy} &Unlimited& $\times$ &\checkmark&\\
\midrule
\bottomrule[1pt]
\end{tabular}
 \captionof{subtable}{Comparison of the different types of homomorphic encryption.}\label{tab5}
    \centering
    
\begin{tabular}{c|c|c|C{11.8cm}}
\toprule[1pt]
\midrule
\textbf{Schemes} & \textbf{Operation} & \textbf{Method} & \textbf{Technique \& Remark} \\ \midrule
Liu \textit{et al.}\cite{liu2019secure} & Paillier & PHE & Leveraging the additive properties of PHE to protect models from exposure to the fusion server. \\ \midrule
BatchCrypt\cite{zhang2020batchcrypt} & Paillier & PHE & Batch process multiple gradients into a large integer and encrypt it with HE. \\ \midrule
FLashe \cite{jiang2021flashe} & Paillier & PHE  &Using symmetric keys to accelerate HE encryption and decryption.     \\ \midrule
Li \textit{et al.} \cite{li2024secure}  & Paillier & PHE & Implement a dual-server model, leveraging a proxy re-encryption scheme with additive homomorphic properties. \\ \midrule
Kumar \textit{et al.} \cite{kumar2024revamping} & CKKS & FHE & Federated Cryptography Defense, implementing FHE through row-based cryptographic transformation.   \\ \midrule
Poseidon \cite{sav2021poseidon} & Paillier &  FHE &Lattice-based HE enables Single Instruction, Multiple Data (SIMD) operations on encrypted data  \\ \midrule
Rieyan \textit{et al.} \cite{rieyan2024advanced} &Paillier & PHE &Storage and analysis of medical image privacy via data fabric architecture. \\ \midrule
Han \textit{et al.} \cite{han2023adaptive}  &Paillier & PHE & Adaptive batch encryption considers both model privacy and heterogeneity issues. \\ \midrule
Hijazi \textit{et al.} \cite{hijazi2023secure} &CKKS &FHE & Addressing privacy issues in IoT-enabled smart cities by combining FHE and FL. \\ \midrule
Tian \textit{et al.} \cite{tian2024lattice} &FV &PHE &Lattice-based distributed threshold AHE uses secure aggregation protocols to defend against semi-trusted adversaries. \\ 
\midrule
\bottomrule[1pt]
\end{tabular}
 \captionof{subtable}{Comparison of different HE-FL schemes.}\label{tab6}
\end{table*}

Study \cite{phong2017privacy} introduces Additive Homomorphic Encryption (AHE) in the FL systems to protect the privacy of all participants. Specifically, all clients agree on a key pair, with the public key used for encrypting gradients before transfer and the private key used for decrypting the received gradients from the server. Similarly, study \cite{hao2019towards} follows the AHE scheme but leverages PPDM \cite{zhou2015ppdm} for encrypting to mitigate the efficiency bottleneck. Although these works provide privacy protection against a curious server, they face several main limitations. First, the proposed schemes are not robust against collusion between the server and even a single participant. Second, the encryption process incurs a large amount of computational overhead, which cannot be afforded by most lightweight edge devices.

To address the collusion issues, HE-based multi-key aggregation protocol (MKAgg) \cite{li2024secure} has been proposed. The protocol is built on a dual-server model and uses proxy re-encryption technology to convert ciphertexts of different public keys into ciphertexts under a public key. From a multiple encryption perspective, PEFL \cite{hao2019efficient} embeds the homomorphic ciphertext of private gradients into augmented learning with errors (A-LWE) term to defend against attacks performed through collusion. By employing multiparty lattice-based cryptography, POSEIDON \cite{sav2021poseidon} further increases the tolerance to collusion among up to $N-1$ parties.

To reduce the formidable overhead associated with encrypting all gradient parameters, MASKCRYPT \cite{hu2024maskcrypt} has been proposed. Instead of encrypting entire model updates, MASKCRYPT applies an encryption mask to select a small portion of the updates for encryption. Compared to encrypting the entire model updates, MASKCRYPT achieves a 4.15× reduction in communication overhead. A recent study, Federated Cryptography Defense \cite{kumar2024revamping}, introduces KD in HE to enhance robustness against both evasion black-box data poisoning and model inversion attacks. Instead of gradient space, FCD transfers the entire learning process into an encrypted data space and leverages a Kullback-Leibler (KL) divergence-based distillation loss to support the KD process. 

The huge communication and encryption overhead are long-term bottlenecks for HE-FL schemes. Today, many studies have proposed methods to reduce these encryption and communication costs and to mitigate this research gap. We summarize these studies in Table \ref{tab6}.

\subsection{Secure Multi-Party Computation (SMPC)}
SMPC was first proposed by Yao \cite{yao1982protocols} to address the problem known as the "two millionaires problem," where two millionaires are interested in knowing which of them is richer without revealing their actual wealth. In the FL context, SMPC provides potential solutions to the problem of securely computing an agreed-upon function without a trusted server, ensuring that each participant learns only the final computation result without gaining any additional information about others.

To achieve SMPC, the study \cite{bonawitz2017practical} proposed secure aggregated FL (namely PPML) based on a Secret Sharing and Key Agreement Protocol. Under PPML, clients' gradients are divided into multiple encrypted shares and then aggregated once a preset number of shares are collected. Since the proposed PPML does not require the gradients of all clients, it can potentially benefit FL by tolerating stragglers. The sharing ($SS.share (\cdot)$)and reconstruction ($SS.recon (\cdot)$) processes of secret sharing shown as following, where $S$ represents the secret, $u$ is the client that belongs to all participants $U$ and $t$ is the threshold for reconstructing $S$:

\begin{equation}
    SS.share(s,t,U)\rightarrow S_u, u\in U 
\end{equation}
\begin{equation}
    SS.recon\{(s_u), u\in U,t\}\rightarrow S 
\end{equation}

MLGuard \cite{khazbak2020mlguard} further enhances the robustness of secret sharing-based FL to poisoning attacks. In each iteration, clients split their local model updates into two parts using secret sharing and send them to two servers separately. After excluding suspicious gradients with low cosine similarity, each server aggregates the shares received and then collaborates with the other server to generate the global model. Considering resource-constrained edge devices, the study \cite{chen2024secure} proposes an efficient and secure SMPC-FL algorithm. This algorithm uses compressed sensing technology to compress transmitted data and introduces additive secret sharing (ASS) and functional encryption (FE) to construct a new inner product functional encryption for executing the SMPC-FL algorithm (FL-IPFE).

While multiple SMPC-FL studies have demonstrated effectiveness in privacy protection, most of these works incur significant computation and communication costs. Effectively deploying SMPC in network- / device-heterogeneous FL scenarios remains an open research question.

\subsection{Trusted Execution Environment (TEE)} TEE has been proposed to protect the security of data and computing in untrusted remote environments. By creating a secure area through the isolation of certain software and hardware resources, TEE ensures that the confidentiality, integrity, and measurability of programs and data running within it are protected from external interference.

Under FL architectures, TEE allows local models and the aggregation process to be loaded in a trusted and private environment (e.g., Intel SGX \cite{cryptoeprint:2016/086}), which mitigates the information leakage and potentially defend against gradient crafting (i.e., model poisoning attacks) and training skipping (i.e., free-riding attacks), mitigating the information leakage \cite{mo2021ppfl, chen2020training}. However, besides suffering from side-channel attacks, current secure enclaves are also limited in memory and provide access only to CPU resources, which conflicts with the GPU computing requirements needed for large ML models and complex learning tasks. Thus, it remains an open question how to effectively deploy FL across secure enclaves, edge computing resources, and client devices.

\section{Challenges and Future Directions}\label{SEC7}
Based on the threats and defenses discussed above, we summarize the effectiveness of various defense methods against different FL threats in Table \ref{summary}. It can be observed that most existing defense solutions are effective against limited threats and fail to provide comprehensive protection for both the utility and privacy of FL. In this section, we remark on these threats and defenses, discuss their opportunities and bottlenecks, and explore future research trends for the trustworthy FL community.

\newcolumntype{C}[1]{>{\centering\arraybackslash}m{#1}}
\begin{table*}[htbp]
\caption{Illustration of the effectiveness of various defense methods against different FL threats.}
\begin{center}
\begin{tabular}{c|C{1.3cm}|C{1.3cm}|C{1.3cm}|C{1.3cm}|C{1.3cm}|C{1.3cm}|C{1.3cm}|c}
\toprule[1pt]
\midrule
\diagbox{Defends \ }{Threats}&Poisoning&Promise Break&Sybil&Communi &Evasion &Inference &Inversion& Eavesdrop \\
\midrule
Data Sanitization&\halfcirc \ \fullcirc &\emptycirc &\emptycirc &\emptycirc &\fullcirc &\emptycirc &\emptycirc &\emptycirc \\\midrule
Anomaly Detection&\halfcirc &\halfcirc &\emptycirc &\emptycirc &\emptycirc &\emptycirc &\emptycirc &\emptycirc\\\midrule
Adversarial Training&\halfcirc &\emptycirc &\emptycirc &\emptycirc &\fullcirc &\emptycirc &\emptycirc &\emptycirc \\\midrule
Byzantine-R Aggregation&\halfcirc \ \fullcirc &\emptycirc &\halfcirc \ \fullcirc &\emptycirc &\emptycirc &\emptycirc &\emptycirc&\emptycirc\\\midrule
Model Compression&\halfcirc &\emptycirc &\emptycirc &\halfcirc &\halfcirc &\halfcirc &\halfcirc &\emptycirc \\\midrule
Trusted Execution E&\halfcirc &\halfcirc &\emptycirc &\fullcirc &\emptycirc &\halfcirc &\halfcirc &\fullcirc\\\midrule
Differential Privacy&\halfcirc &\emptycirc &\emptycirc &\emptycirc &\fullcirc &\fullcirc &\fullcirc &\emptycirc\\\midrule
Homomorphic Encryption&\emptycirc &\emptycirc &\emptycirc &\emptycirc &\emptycirc &\fullcirc &\emptycirc &\fullcirc\\\midrule
Blockchain&\emptycirc \ \halfcirc& \halfcirc&\fullcirc &\emptycirc  &\emptycirc &\halfcirc \ \fullcirc & \fullcirc &\halfcirc\\\midrule
Decentralized FL&\emptycirc &\emptycirc &\emptycirc &\emptycirc &\emptycirc &\halfcirc &\halfcirc &\emptycirc\\\midrule
Secure Multi-parity Comp&\emptycirc &\emptycirc &\emptycirc &\emptycirc &\emptycirc &\halfcirc \ \fullcirc &\emptycirc &\emptycirc\\\midrule
\multicolumn{9}{c}{$\fullcirc \ \halfcirc \ \emptycirc$ represent the method that can effectively, partially, or can not defend against the target attacks. \  Based on different capacities of the adversary,}\\
\multicolumn{9}{c}{the defend methods could show different performances; thus, we use two marks to represent a range of its robustness.}\\
\bottomrule[1pt]
\end{tabular}
\label{summary}
\end{center}
\end{table*}

\subsection{Large FL Attack Surface Reduction}
To effectively degrade model performance and avoid detection, malicious clients must introduce sufficient perturbation to their model updates while ensuring that the crafted gradients remain similar to benign ones \cite{fang2020local}. Essentially, the difference between benign gradients can be considered a surface within which attackers can craft their gradients without being detected. A large attack surface provides opportunities for attacks and poses challenges for defenders \cite{li2023enhancing}. In FL, this large attack surface primarily stems from two factors: (i) the high dimensionality of the model and (ii) the non-IID data distribution among participants. Therefore, reducing the FL attack surface by addressing these two aspects could be a promising direction for future research.

\subsection{New Perspective for Attacks}
FL involves multiple participants collaborating to train a single joint model, making fairness across participants an increasingly important focus in recent studies \cite{chen2023privacy}. These fairness enhancement solutions aim to guarantee similar representation across different data distributions, ensure that different participants have an equal probability of being selected, and provide reasonable rewards. However, most of these solutions assume the honesty and trustworthiness of participants, which may not hold in real-world scenarios, thereby creating new opportunities for attacks. Given that fairness-oriented approaches are still in their infancy, it is worthwhile to explore their vulnerabilities and develop corresponding defense solutions.

\subsection{Attackers vs Disadvantaged Clients}
Due to the non-IID nature, FL participants are usually heterogeneous in terms of computing capacity, data quality, data distribution, and bandwidth. For instance, a UAV deployed in a high-altitude area may capture more unique photos compared to a UAV in urban environments. Despite acting honestly, it may be mistakenly identified as malicious if its gradients exhibit low evaluation performance or deviate significantly from others. This misidentification occurs because current Byzantine-resilient aggregation methods identify malicious clients based on the performance or statistical values of their gradients. Notably, unusual data distributions often carry valuable information for model training. Therefore, it is important to study how to distinguish disadvantaged clients from attackers and ensure their participation.

\subsection{Federated Unlearning}
Current FL defense methods emphasize analyzing clients' behavior or gradient performance in the current learning iteration to evaluate their honesty. However, many attacks can achieve their objectives (e.g., introducing a backdoor) over several learning rounds. In such cases, excluding a suspicious client from only the current learning iteration may still allow the introduced perturbations to remain in the joint model. Federated unlearning \cite{wu2022federated} has emerged as a new technology, originally designed to eliminate the contributions of a client once they leave the collaboration. By introducing federated unlearning, defenses are expected to resolve the issue of residual poisoning in the model.

\subsection{Research on VFL, FTL, and DFL}
Existing research on trustworthy FL mainly focuses on the HFL setting. However, since VFL, FTL, and HFL have different shared feature spaces, defense methods for HFL may not be directly applicable to VFL and FTL systems, and their effectiveness remains unclear. Furthermore, due to their different label spaces, FTL and TFL also present specific attack opportunities (e.g., overall feature space exposure and sample ID leakage). Overall, there remains a significant gap in research for VFL and FTL. 

Furthermore, as an emerging learning paradigm, the research on DFL is still in its infancy. Without a centralized node, traditional byzantine-robust aggregation methods are potentially inapplicable. Enhancing the robustness of DFL requires further investigation.

\subsection{Emerging Privacy Requirements}
Traditional privacy requirements in FL focus on the information at the training data level (e.g., original data samples) or client level (e.g., participation). However, with the evolution of machine learning, current models have become much larger, requiring extensive costs for training (consider the server distributing incentives to clients). Additionally, model architectures are often well-tuned to achieve high learning performance. Consequently, well-trained models and elaborate-tuned model architectures have become new objectives for privacy protection \cite{kalra2023decentralized}, which servers may be reluctant to share during training. Therefore, it is essential to explore new privacy requirements and design new FL algorithms that fulfill these emerging requirements.

\subsection{Both Privacy- and Security- Enhanced FL algorithms}
Although various FL algorithms have been proposed, most of these methods focus on single issues and fail to address both privacy and security concerns simultaneously. Unfortunately, directly combining different techniques may not result in a cumulative effect and even introduce new attack opportunities. For instance, DP has been shown to potentially enlarge the attack surface \cite{li2023enhancing}, which may weaken the effectiveness of statistical-feature-based Byzantine-robust aggregation methods. This area remains largely unexplored, and there are significant gaps in achieving FL algorithms that enhance both privacy and security.

\section{Conclusion}\label{SEC8}
FL is an innovative ML paradigm that allows multiple clients to collaboratively train a model without sharing their raw data. Due to its inherent privacy-preserving nature, FL has the potential to safeguard data privacy, effectively transforming data from "visible and usable" to "invisible yet usable." This capability extends the applicability of data-driven ML models to real-world scenarios. In this paper, we comprehensively survey the most significant and cutting-edge threats and defense mechanisms throughout the entire FL lifecycle. Our detailed discussion and comparison enhance the understanding of the attack and defense dynamics in FL. In the section on future research trends, we highlight areas within FL that necessitate thorough research and exploration. This survey paper is an invaluable resource for researchers, designers, and engineers engaged in FL security. It offers insights into the current challenges and future directions for research in this evolving field.
\bibliographystyle{IEEEtran}
\bibliography{paper}

\begin{thebibliography}{100}
\providecommand{\url}[1]{#1}
\csname url@samestyle\endcsname
\providecommand{\newblock}{\relax}
\providecommand{\bibinfo}[2]{#2}
\providecommand{\BIBentrySTDinterwordspacing}{\spaceskip=0pt\relax}
\providecommand{\BIBentryALTinterwordstretchfactor}{4}
\providecommand{\BIBentryALTinterwordspacing}{\spaceskip=\fontdimen2\font plus
\BIBentryALTinterwordstretchfactor\fontdimen3\font minus \fontdimen4\font\relax}
\providecommand{\BIBforeignlanguage}[2]{{%
\expandafter\ifx\csname l@#1\endcsname\relax
\typeout{** WARNING: IEEEtran.bst: No hyphenation pattern has been}%
\typeout{** loaded for the language `#1'. Using the pattern for}%
\typeout{** the default language instead.}%
\else
\language=\csname l@#1\endcsname
\fi
#2}}
\providecommand{\BIBdecl}{\relax}
\BIBdecl

\bibitem{li2021survey}
Q.~Li, Z.~Wen, Z.~Wu, S.~Hu, N.~Wang, Y.~Li, X.~Liu, and B.~He, ``A survey on federated learning systems: vision, hype and reality for data privacy and protection,'' \emph{IEEE Transactions on Knowledge and Data Engineering}, vol.~35, no.~1, pp. 3347--3366, 2021.

\bibitem{gabrielli2023survey}
H.~Ye, L.~Liang, and G.~Y. Li, ``Decentralized federated learning with unreliable communications,'' \emph{IEEE Journal of Selected Topics in Signal Processing}, vol.~16, no.~3, pp. 487--500, 2022.

\bibitem{kairouz2021advances}
P.~Kairouz, H.~B. McMahan, B.~Avent, A.~Bellet, M.~Bennis, A.~N. Bhagoji, K.~Bonawitz, Z.~Charles, G.~Cormode, R.~Cummings \emph{et~al.}, ``Advances and open problems in federated learning,'' \emph{Foundations and Trends{\textregistered} in Machine Learning}, vol.~14, no. 1--2, pp. 1--210, 2021.

\bibitem{mcmahan2017communication}
B.~McMahan, E.~Moore, D.~Ramage, S.~Hampson, and B.~A. y~Arcas, ``Communication-efficient learning of deep networks from decentralized data,'' in \emph{Artificial intelligence and statistics (AISTATS)}.\hskip 1em plus 0.5em minus 0.4em\relax PMLR, 2017, pp. 1273--1282.

\bibitem{hard2018federated}
A.~Hard, K.~Rao, R.~Mathews, S.~Ramaswamy, F.~Beaufays, S.~Augenstein, H.~Eichner, C.~Kiddon, and D.~Ramage, ``Federated learning for mobile keyboard prediction,'' \emph{arXiv preprint arXiv:1811.03604}, 2018.

\bibitem{apple}
Apple, ``Designing for privacy (video and slide deck), wwdc,'' https://developer.apple.com/videos/play/wwdc2019/708, 2019.

\bibitem{WeBank}
WeBank, ``Webank and swiss re signed cooperation, mou,'' https://finance.yahoo.com/news/webank-swiss-signed-cooperation-mou-112300218.html, 2019.

\bibitem{li2024secure}
Y.~Li, J.~Lai, R.~Zhang, and M.~Sun, ``Secure and efficient multi-key aggregation for federated learning,'' \emph{Information Sciences}, vol. 654, p. 119830, 2024.

\bibitem{damaskinos2019aggregathor}
G.~Damaskinos, E.-M. El-Mhamdi, R.~Guerraoui, A.~Guirguis, and S.~Rouault, ``Aggregathor: Byzantine machine learning via robust gradient aggregation,'' in \emph{Machine Learning and Systems (MLSys)}.\hskip 1em plus 0.5em minus 0.4em\relax ACM, 2019, pp. 81--106.

\bibitem{el2022genuinely}
E.-M. El-Mhamdi, R.~Guerraoui, A.~Guirguis, L.-N. Hoang, and S.~Rouault, ``Genuinely distributed byzantine machine learning,'' \emph{Distributed Computing}, vol.~35, pp. 305--331, 2022.

\bibitem{fang2020local}
M.~Fang, X.~Cao, J.~Jia, and N.~Gong, ``Local model poisoning attacks to byzantine-robust federated learning,'' in \emph{USENIX Security Symposium (USENIX Security)}.\hskip 1em plus 0.5em minus 0.4em\relax USENIX Association, 2020, pp. 1605--1622.

\bibitem{9806416}
X.~Gong, Y.~Chen, Q.~Wang, and W.~Kong, ``Backdoor attacks and defenses in federated learning: State-of-the-art, taxonomy, and future directions,'' \emph{IEEE Wireless Communications}, vol.~30, no.~2, pp. 114--121, 2023.

\bibitem{li2022backdoor}
Y.~Li, Y.~Jiang, Z.~Li, and S.-T. Xia, ``Backdoor learning: A survey,'' \emph{IEEE Transactions on Neural Networks and Learning Systems}, vol.~1, no.~1, pp. 1--18, 2022.

\bibitem{baghbani2022application}
A.~Baghbani, T.~Choudhury, S.~Costa, and J.~Reiner, ``Application of artificial intelligence in geotechnical engineering: A state-of-the-art review,'' \emph{Earth-Science Reviews}, vol. 228, no.~1, pp. 1--26, 2022.

\bibitem{shokri2017membership}
R.~Shokri, M.~Stronati, C.~Song, and V.~Shmatikov, ``Membership inference attacks against machine learning models,'' in \emph{2017 IEEE symposium on security and privacy (SP)}.\hskip 1em plus 0.5em minus 0.4em\relax IEEE, 2017, pp. 3--18.

\bibitem{nasr2019comprehensive}
M.~Nasr, R.~Shokri, and A.~Houmansadr, ``Comprehensive privacy analysis of deep learning: Passive and active white-box inference attacks against centralized and federated learning,'' in \emph{2019 IEEE symposium on security and privacy (SP)}.\hskip 1em plus 0.5em minus 0.4em\relax IEEE, 2019, pp. 739--753.

\bibitem{zhang2023privacy}
S.~Zhang and J.~Zhu, ``Privacy protection federated learning framework based on blockchain and committee consensus in iot devices,'' in \emph{2023 IEEE 47th Annual Computers, Software, and Applications Conference (COMPSAC)}.\hskip 1em plus 0.5em minus 0.4em\relax IEEE, 2023, pp. 627--636.

\bibitem{ozdayi2021defending}
M.~S. Ozdayi, M.~Kantarcioglu, and Y.~R. Gel, ``Defending against backdoors in federated learning with robust learning rate,'' in \emph{Proceedings of the AAAI Conference on Artificial Intelligence}, vol.~35, no.~10, 2021, pp. 9268--9276.

\bibitem{taheri2020defending}
R.~Taheri, R.~Javidan, M.~Shojafar, Z.~Pooranian, A.~Miri, and M.~Conti, ``On defending against label flipping attacks on malware detection systems,'' \emph{Neural Computing and Applications}, vol.~32, pp. 14\,781--14\,800, 2020.

\bibitem{cao2020fltrust}
X.~Cao, M.~Fang, J.~Liu, and N.~Z. Gong, ``Fltrust: Byzantine-robust federated learning via trust bootstrapping,'' in \emph{ISOC Network and Distributed System Security Symposium (NDSS)}.\hskip 1em plus 0.5em minus 0.4em\relax ISOC, 2021, pp. 1--18.

\bibitem{xia2023poisoning}
G.~Xia, J.~Chen, C.~Yu, and J.~Ma, ``Poisoning attacks in federated learning: A survey,'' \emph{IEEE Access}, vol.~11, pp. 10\,708--10\,722, 2023.

\bibitem{wen2023survey}
J.~Wen, Z.~Zhang, Y.~Lan, Z.~Cui, J.~Cai, and W.~Zhang, ``A survey on federated learning: challenges and applications,'' \emph{International Journal of Machine Learning and Cybernetics}, vol.~14, no.~2, pp. 513--535, 2023.

\bibitem{mothukuri2021survey}
V.~Mothukuri, R.~M. Parizi, S.~Pouriyeh, Y.~Huang, A.~Dehghantanha, and G.~Srivastava, ``A survey on security and privacy of federated learning,'' \emph{Future Generation Computer Systems}, vol. 115, pp. 619--640, 2021.

\bibitem{Gao2022survey}
Y.~Gao, X.~Chen, Y.~Zhang, W.~Wang, H.~Deng, P.~Duan, and P.~Chen, ``A survey of attack and defense techniques for federated learning systems,'' \emph{Chinese Journal Of Computes}, pp. 1--25, 2023.

\bibitem{guo2024white}
Z.~Guo, W.~Li, Y.~Qian, O.~Arandjelovic, and L.~Fang, ``A white-box false positive adversarial attack method on contrastive loss based offline handwritten signature verification models,'' in \emph{International Conference on Artificial Intelligence and Statistics}.\hskip 1em plus 0.5em minus 0.4em\relax PMLR, 2024, pp. 901--909.

\bibitem{kumar2023impact}
K.~N. Kumar, C.~K. Mohan, and L.~R. Cenkeramaddi, ``The impact of adversarial attacks on federated learning: A survey,'' \emph{IEEE Transactions on Pattern Analysis and Machine Intelligence}, 2023.

\bibitem{zhou2021deep}
X.~Zhou, M.~Xu, Y.~Wu, and N.~Zheng, ``Deep model poisoning attack on federated learning,'' \emph{Future Internet}, vol.~13, no.~3, pp. 1--14, 2021.

\bibitem{10121613}
K.~Wei, J.~Li, M.~Ding, C.~Ma, Y.-S. Jeon, and H.~Vincent~Poor, ``Covert model poisoning against federated learning: Algorithm design and optimization,'' \emph{IEEE Transactions on Dependable and Secure Computing}, pp. 1--14, 2023.

\bibitem{xiao2012adversarial}
H.~Xiao, H.~Xiao, and C.~Eckert, ``Adversarial label flips attack on support vector machines,'' in \emph{European Conference on Artificial Intelligence (ECAI)}.\hskip 1em plus 0.5em minus 0.4em\relax IOS Press, 2012, pp. 870--875.

\bibitem{LI2024103936}
X.~Li, N.~Wang, S.~Yuan, and Z.~Guan, ``Fedimp: Parameter importance-based model poisoning attack against federated learning system,'' \emph{Computers \& Security}, p. 103936, 2024.

\bibitem{baruch2019little}
G.~Baruch, M.~Baruch, and Y.~Goldberg, ``A little is enough: Circumventing defenses for distributed learning,'' in \emph{Advances in Neural Information Processing Systems (NeurIPS)}.\hskip 1em plus 0.5em minus 0.4em\relax MIT Press, 2019, pp. 1--11.

\bibitem{xie2020fall}
C.~Xie, O.~Koyejo, and I.~Gupta, ``Fall of empires: Breaking byzantine-tolerant sgd by inner product manipulation,'' in \emph{Uncertainty in Artificial Intelligence (UAI)}.\hskip 1em plus 0.5em minus 0.4em\relax PMLR, 2020, pp. 261--270.

\bibitem{blanchard2017machine}
P.~Blanchard, E.~M. El~Mhamdi, R.~Guerraoui, and J.~Stainer, ``Machine learning with adversaries: Byzantine tolerant gradient descent,'' in \emph{Advances in Neural Information Processing Systems (NeurIPS)}.\hskip 1em plus 0.5em minus 0.4em\relax MIT Press, 2017, pp. 1--11.

\bibitem{yin2018byzantine}
D.~Yin, Y.~Chen, R.~Kannan, and P.~Bartlett, ``Byzantine-robust distributed learning: Towards optimal statistical rates,'' in \emph{International Conference on Machine Learning (ICML)}.\hskip 1em plus 0.5em minus 0.4em\relax PMLR, 2018, pp. 5650--5659.

\bibitem{karimireddy2021byzantine}
S.~P. Karimireddy, L.~He, and M.~Jaggi, ``Byzantine-robust learning on heterogeneous datasets via bucketing,'' in \emph{International Conference on Learning Representations (ICLR)}, 2022.

\bibitem{shejwalkar2021manipulating}
V.~Shejwalkar and A.~Houmansadr, ``Manipulating the byzantine: Optimizing model poisoning attacks and defenses for federated learning,'' in \emph{NDSS}, 2021.

\bibitem{tolpegin2020data}
V.~Tolpegin, S.~Truex, M.~E. Gursoy, and L.~Liu, ``Data poisoning attacks against federated learning systems,'' in \emph{European Symposium on Research in Computer Security (ESORICS)}.\hskip 1em plus 0.5em minus 0.4em\relax Springer, 2020, pp. 480--501.

\bibitem{lewis2023attacks}
C.~Lewis, V.~Varadharajan, and N.~Noman, ``Attacks against federated learning defense systems and their mitigation,'' \emph{Journal of Machine Learning Research}, vol.~24, no.~30, pp. 1--50, 2023.

\bibitem{zhang2021label}
H.~Zhang, N.~Cheng, Y.~Zhang, and Z.~Li, ``Label flipping attacks against naive bayes on spam filtering systems,'' \emph{Applied Intelligence}, vol.~51, pp. 4503--4514, 2021.

\bibitem{sharma2022catboost}
R.~Sharma, G.~Sharma, and M.~Pattanaik, ``A catboost based approach to detect label flipping poisoning attack in hardware trojan detection systems,'' \emph{Journal of Electronic Testing}, vol.~38, no.~6, pp. 667--682, 2022.

\bibitem{nguyen2020poisoning}
T.~D. Nguyen, P.~Rieger, M.~Miettinen, and A.-R. Sadeghi, ``Poisoning attacks on federated learning-based iot intrusion detection system,'' in \emph{Workshop on Decentralized IoT Systems and Security (DISS)}.\hskip 1em plus 0.5em minus 0.4em\relax ISOC, 2020, pp. 1--7.

\bibitem{bagdasaryan2020backdoor}
E.~Bagdasaryan, A.~Veit, Y.~Hua, D.~Estrin, and V.~Shmatikov, ``How to backdoor federated learning,'' in \emph{International Conference on Artificial Intelligence and Statistics (AISTATS)}.\hskip 1em plus 0.5em minus 0.4em\relax PMLR, 2020, pp. 2938--2948.

\bibitem{wang2020attack}
H.~Wang, K.~Sreenivasan, S.~Rajput, H.~Vishwakarma, S.~Agarwal, J.-y. Sohn, K.~Lee, and D.~Papailiopoulos, ``Attack of the tails: Yes, you really can backdoor federated learning,'' in \emph{Advances in Neural Information Processing Systems (NeurIPS)}.\hskip 1em plus 0.5em minus 0.4em\relax MIT Press, 2020, pp. 16\,070--16\,084.

\bibitem{wan2021shielding}
W.~Wan, J.~Lu, S.~Hu, L.~Y. Zhang, and X.~Pei, ``Shielding federated learning: A new attack approach and its defense,'' in \emph{IEEE Wireless Communications and Networking Conference (WCNC)}.\hskip 1em plus 0.5em minus 0.4em\relax IEEE, 2021, pp. 1--7.

\bibitem{fraboni2021free}
Y.~Fraboni, R.~Vidal, and M.~Lorenzi, ``Free-rider attacks on model aggregation in federated learning,'' in \emph{International Conference on Artificial Intelligence and Statistics (AISTATS)}.\hskip 1em plus 0.5em minus 0.4em\relax PMLR, 2021, pp. 1846--1854.

\bibitem{lin2019free}
B.~Pej{\'o} and G.~Bicz{\'o}k, ``Quality inference in federated learning with secure aggregation,'' \emph{IEEE Transactions on Big Data}, vol.~9, no.~5, pp. 1430--1438, 2023.

\bibitem{liu2020incentives}
Y.~Liu and J.~Wei, ``Incentives for federated learning: a hypothesis elicitation approach,'' \emph{arXiv preprint arXiv:2007.10596}, 2020.

\bibitem{zeng2021comprehensive}
R.~Zeng, C.~Zeng, X.~Wang, B.~Li, and X.~Chu, ``A comprehensive survey of incentive mechanism for federated learning,'' \emph{arXiv preprint arXiv:2106.15406}, 2021.

\bibitem{weng2019deepchain}
J.~Weng, J.~Weng, J.~Zhang, M.~Li, Y.~Zhang, and W.~Luo, ``Deepchain: Auditable and privacy-preserving deep learning with blockchain-based incentive,'' \emph{IEEE Transactions on Dependable and Secure Computing}, vol.~18, no.~5, pp. 2438--2455, 2019.

\bibitem{fung2018mitigating}
T.~Tuor, S.~Wang, B.~J. Ko, C.~Liu, and K.~K. Leung, ``Overcoming noisy and irrelevant data in federated learning,'' in \emph{International Conference on Pattern Recognition (ICPR)}.\hskip 1em plus 0.5em minus 0.4em\relax IEEE, 2021, pp. 5020--5027.

\bibitem{fung2020limitations}
X.~Cao, Z.~Zhang, J.~Jia, and N.~Z. Gong, ``Flcert: Provably secure federated learning against poisoning attacks,'' \emph{IEEE Transactions on Information Forensics and Security}, vol.~17, pp. 3691--3705, 2022.

\bibitem{LI2024120475}
Y.~Li, W.~Ding, H.~Chen, W.~Bao, and D.~Yuan, ``Contribution-wise byzantine-robust aggregation for class-balanced federated learning,'' \emph{Information Sciences}, vol. 667, p. 120475, 2024.

\bibitem{agrawal2018detection}
S.~Agrawal, M.~L. Das, and J.~Lopez, ``Detection of node capture attack in wireless sensor networks,'' \emph{IEEE Systems Journal}, vol.~13, no.~1, pp. 238--247, 2018.

\bibitem{9994772}
A.~Vangala, A.~K. Das, A.~Mitra, S.~K. Das, and Y.~Park, ``Blockchain-enabled authenticated key agreement scheme for mobile vehicles-assisted precision agricultural iot networks,'' \emph{IEEE Transactions on Information Forensics and Security}, vol.~18, pp. 904--919, 2023.

\bibitem{yao2018two}
X.~Yao, C.~Huang, and L.~Sun, ``Two-stream federated learning: Reduce the communication costs,'' in \emph{IEEE Visual Communications and Image Processing (VCIP)}.\hskip 1em plus 0.5em minus 0.4em\relax IEEE, 2018, pp. 1--4.

\bibitem{goodfellow2014explaining}
I.~J. Goodfellow, J.~Shlens, and C.~Szegedy, ``Explaining and harnessing adversarial examples,'' in \emph{International Conference on Learning Representations (ICLR)}.\hskip 1em plus 0.5em minus 0.4em\relax IMLS, 2015, pp. 1--11.

\bibitem{szegedy2015going}
C.~Szegedy, W.~Liu, Y.~Jia, P.~Sermanet, S.~Reed, D.~Anguelov, D.~Erhan, V.~Vanhoucke, and A.~Rabinovich, ``Going deeper with convolutions,'' in \emph{IEEE Conference on Computer Vision and Pattern Recognition (CVPR)}.\hskip 1em plus 0.5em minus 0.4em\relax IEEE, 2015, pp. 1--9.

\bibitem{madry2018towards}
A.~Madry, A.~Makelov, L.~Schmidt, D.~Tsipras, and A.~Vladu, ``Towards deep learning models resistant to adversarial attacks,'' in \emph{International Conference on Learning Representations (ICLR)}.\hskip 1em plus 0.5em minus 0.4em\relax IMLS, 2018, pp. 1--23.

\bibitem{kurakin2017adversarial}
A.~Kurakin, I.~J. Goodfellow, and S.~Bengio, ``Adversarial machine learning at scale,'' in \emph{International Conference on Learning Representations (ICLR)}.\hskip 1em plus 0.5em minus 0.4em\relax IMLS, 2017, pp. 1--17.

\bibitem{chen2017zoo}
P.-Y. Chen, H.~Zhang, Y.~Sharma, J.~Yi, and C.-J. Hsieh, ``Zoo: Zeroth order optimization based black-box attacks to deep neural networks without training substitute models,'' in \emph{ACM Workshop on Artificial Intelligence and Security (AISec)}.\hskip 1em plus 0.5em minus 0.4em\relax ACM, 2017, pp. 15--26.

\bibitem{brendel2018decision}
W.~Brendel, J.~Rauber, and M.~Bethge, ``Decision-based adversarial attacks: Reliable attacks against black-box machine learning models,'' in \emph{International Conference on Learning Representations (ICLR)}.\hskip 1em plus 0.5em minus 0.4em\relax IMLS, 2018, pp. 1--17.

\bibitem{pang2022attacking}
Q.~Pang, Y.~Yuan, S.~Wang, and W.~Zheng, ``Adi: Adversarial dominating inputs in vertical federated learning systems,'' in \emph{2023 IEEE Symposium on Security and Privacy (SP)}.\hskip 1em plus 0.5em minus 0.4em\relax IEEE, 2023, pp. 1875--1892.

\bibitem{deng2009imagenet}
J.~Deng, W.~Dong, R.~Socher, L.-J. Li, K.~Li, and L.~Fei-Fei, ``Imagenet: A large-scale hierarchical image database,'' in \emph{2009 IEEE conference on computer vision and pattern recognition}.\hskip 1em plus 0.5em minus 0.4em\relax Ieee, 2009, pp. 248--255.

\bibitem{zhu2019deep}
L.~Zhu, Z.~Liu, and S.~Han, ``Deep leakage from gradients,'' \emph{Advances in neural information processing systems}, vol.~32, 2019.

\bibitem{geiping2020inverting}
J.~Geiping, H.~Bauermeister, H.~Dr{\"o}ge, and M.~Moeller, ``Inverting gradients-how easy is it to break privacy in federated learning?'' \emph{Advances in neural information processing systems}, vol.~33, pp. 16\,937--16\,947, 2020.

\bibitem{yin2020dreaming}
H.~Yin, P.~Molchanov, J.~M. Alvarez, Z.~Li, A.~Mallya, D.~Hoiem, N.~K. Jha, and J.~Kautz, ``Dreaming to distill: Data-free knowledge transfer via deepinversion,'' in \emph{Proceedings of the IEEE/CVF Conference on Computer Vision and Pattern Recognition}, 2020, pp. 8715--8724.

\bibitem{yin2021see}
H.~Yin, A.~Mallya, A.~Vahdat, J.~M. Alvarez, J.~Kautz, and P.~Molchanov, ``See through gradients: Image batch recovery via gradinversion,'' in \emph{Proceedings of the IEEE/CVF Conference on Computer Vision and Pattern Recognition}, 2021, pp. 16\,337--16\,346.

\bibitem{Li_2022_CVPR}
Z.~Li, J.~Zhang, L.~Liu, and J.~Liu, ``Auditing privacy defenses in federated learning via generative gradient leakage,'' in \emph{Proceedings of the IEEE/CVF Conference on Computer Vision and Pattern Recognition (CVPR)}, June 2022, pp. 10\,132--10\,142.

\bibitem{hatamizadeh2022gradvit}
A.~Hatamizadeh, H.~Yin, H.~R. Roth, W.~Li, J.~Kautz, D.~Xu, and P.~Molchanov, ``Gradvit: Gradient inversion of vision transformers,'' in \emph{Proceedings of the IEEE/CVF Conference on Computer Vision and Pattern Recognition (CVPR)}, June 2022, pp. 10\,021--10\,030.

\bibitem{salem2019ml}
A.~Salem, Y.~Zhang, M.~Humbert, P.~Berrang, M.~Fritz, and M.~Backes, ``Ml-leaks: Model and data independent membership inference attacks and defenses on machine learning models,'' in \emph{Proceedings of the 26th Annual Network and Distributed System Security Symposium (NDSS)}, 2019.

\bibitem{li2021membership}
Z.~Li and Y.~Zhang, ``Membership leakage in label-only exposures,'' in \emph{Proceedings of the 2021 ACM SIGSAC Conference on Computer and Communications Security}, 2021, pp. 880--895.

\bibitem{Gu2023Survey}
Y.~Gu and Y.~Bai, ``Survey on security and privacy of federated learning models,'' \emph{Ruan Jian Xue Bao/Journal of Software}, vol.~34, no.~6, pp. 2833--2864, 2023.

\bibitem{melis2019exploiting}
L.~Melis, C.~Song, E.~De~Cristofaro, and V.~Shmatikov, ``Exploiting unintended feature leakage in collaborative learning,'' in \emph{2019 IEEE symposium on security and privacy (SP)}.\hskip 1em plus 0.5em minus 0.4em\relax IEEE, 2019, pp. 691--706.

\bibitem{ganju2018property}
K.~Ganju, Q.~Wang, W.~Yang, C.~A. Gunter, and N.~Borisov, ``Property inference attacks on fully connected neural networks using permutation invariant representations,'' in \emph{Proceedings of the 2018 ACM SIGSAC conference on computer and communications security}, 2018, pp. 619--633.

\bibitem{274683}
W.~Zhang, S.~Tople, and O.~Ohrimenko, ``Leakage of dataset properties in {Multi-Party} machine learning,'' in \emph{30th USENIX Security Symposium (USENIX Security 21)}.\hskip 1em plus 0.5em minus 0.4em\relax USENIX Association, Aug. 2021, pp. 2687--2704.

\bibitem{hitaj2017deep}
B.~Hitaj, G.~Ateniese, and F.~Perez-Cruz, ``Deep models under the gan: information leakage from collaborative deep learning,'' in \emph{Proceedings of the 2017 ACM SIGSAC conference on computer and communications security}, 2017, pp. 603--618.

\bibitem{wang2019beyond}
Z.~Wang, M.~Song, Z.~Zhang, Y.~Song, Q.~Wang, and H.~Qi, ``Beyond inferring class representatives: User-level privacy leakage from federated learning,'' in \emph{IEEE INFOCOM 2019-IEEE conference on computer communications}.\hskip 1em plus 0.5em minus 0.4em\relax IEEE, 2019, pp. 2512--2520.

\bibitem{shokri2015privacy}
R.~Shokri and V.~Shmatikov, ``Privacy-preserving deep learning,'' in \emph{Proceedings of the 22nd ACM SIGSAC conference on computer and communications security}, 2015, pp. 1310--1321.

\bibitem{review2024}
Y.~Sun, Y.~Yan, J.~Cui, G.~Xiong, and J.~Liu, ``Review of deep gradient inversion attacks and defenses in federated learning,'' \emph{Journal of Electronics \& Information Technology}, vol.~46, no.~2, pp. 429--442, 2024.

\bibitem{zou2013eavesdropping}
Y.~Zou, X.~Wang, and W.~Shen, ``Eavesdropping attack in collaborative wireless networks: Security protocols and intercept behavior,'' in \emph{Proceedings of the 2013 IEEE 17th international conference on computer supported cooperative work in design (CSCWD)}.\hskip 1em plus 0.5em minus 0.4em\relax IEEE, 2013, pp. 704--709.

\bibitem{wang2023energy}
T.~Wang, N.~Huang, Y.~Wu, and T.~Q. Quek, ``Energy-efficient wireless federated learning: A secrecy oriented design via sequential artificial jamming,'' \emph{IEEE Transactions on Vehicular Technology}, vol.~72, no.~5, pp. 6412--6427, 2023.

\bibitem{ma2020safeguarding}
C.~Ma, J.~Li, M.~Ding, H.~H. Yang, F.~Shu, T.~Q. Quek, and H.~V. Poor, ``On safeguarding privacy and security in the framework of federated learning,'' \emph{IEEE network}, vol.~34, no.~4, pp. 242--248, 2020.

\bibitem{9839650}
Y.-A. Xie, J.~Kang, D.~Niyato, N.~T.~T. Van, N.~C. Luong, Z.~Liu, and H.~Yu, ``Securing federated learning: A covert communication-based approach,'' \emph{IEEE Network}, vol.~37, no.~1, pp. 118--124, 2023.

\bibitem{xu2021else}
C.~Xu and G.~Neglia, ``What else is leaked when eavesdropping federated learning?'' in \emph{CCS workshop Privacy Preserving Machine Learning (PPML)}, 2021.

\bibitem{10051719}
X.~Hou, J.~Wang, C.~Jiang, X.~Zhang, Y.~Ren, and M.~Debbah, ``Uav-enabled covert federated learning,'' \emph{IEEE Transactions on Wireless Communications}, vol.~22, no.~10, pp. 6793--6809, 2023.

\bibitem{4531146}
G.~F. Cretu, A.~Stavrou, M.~E. Locasto, S.~J. Stolfo, and A.~D. Keromytis, ``Casting out demons: Sanitizing training data for anomaly sensors,'' in \emph{IEEE Symposium on Security and Privacy (SP)}.\hskip 1em plus 0.5em minus 0.4em\relax IEEE, 2008, pp. 81--95.

\bibitem{li2023martfl}
Q.~Li, Z.~Liu, Q.~Li, and K.~Xu, ``martfl: Enabling utility-driven data marketplace with a robust and verifiable federated learning architecture,'' in \emph{Proceedings of the 2023 ACM SIGSAC Conference on Computer and Communications Security}, 2023, pp. 1496--1510.

\bibitem{9248056}
D.~Data, L.~Song, and S.~N. Diggavi, ``Data encoding for byzantine-resilient distributed optimization,'' \emph{IEEE Transactions on Information Theory}, vol.~67, no.~2, pp. 1117--1140, 2021.

\bibitem{li2024feddiv}
J.~Li, G.~Li, H.~Cheng, Z.~Liao, and Y.~Yu, ``Feddiv: Collaborative noise filtering for federated learning with noisy labels,'' in \emph{Proceedings of the AAAI Conference on Artificial Intelligence}, vol.~38, no.~4, 2024, pp. 3118--3126.

\bibitem{koh2022stronger}
P.~W. Koh, J.~Steinhardt, and P.~Liang, ``Stronger data poisoning attacks break data sanitization defenses,'' \emph{Machine Learning}, vol. 111, no.~1, pp. 1--47, 2022.

\bibitem{li2019abnormal}
A.~Yazdinejad, A.~Dehghantanha, R.~M. Parizi, M.~Hammoudeh, H.~Karimipour, and G.~Srivastava, ``Block hunter: Federated learning for cyber threat hunting in blockchain-based iiot networks,'' \emph{IEEE Transactions on Industrial Informatics}, vol.~18, no.~11, pp. 8356--8366, 2022.

\bibitem{meng2021vadaf}
L.~Meng, Y.~Wei, R.~Pan, S.~Zhou, J.~Zhang, and W.~Chen, ``Vadaf: Visualization for abnormal client detection and analysis in federated learning,'' \emph{ACM Transactions on Interactive Intelligent Systems}, vol.~11, no.~4, pp. 1--23, 2021.

\bibitem{hong2023federated}
J.~Hong, H.~Wang, Z.~Wang, and J.~Zhou, ``Federated robustness propagation: Sharing adversarial robustness in heterogeneous federated learning,'' in \emph{AAAI Conference on Artificial Intelligence (AAAI)}.\hskip 1em plus 0.5em minus 0.4em\relax AAAI, 2023, pp. 7893--7901.

\bibitem{li2023enhancing}
Y.~Li, D.~Yuan, A.~S. Sani, and W.~Bao, ``Enhancing federated learning robustness in adversarial environment through clustering non-iid features,'' \emph{Computers \& Security}, vol. 132, no.~1, pp. 1--13, 2023.

\bibitem{li2020learning}
R.~Taheri, M.~Shojafar, M.~Alazab, and R.~Tafazolli, ``Fed-iiot: A robust federated malware detection architecture in industrial iot,'' \emph{IEEE Transactions on Industrial Informatics}, vol.~17, no.~12, pp. 8442--8452, 2020.

\bibitem{chen2018automated}
S.~Chen, M.~Xue, L.~Fan, S.~Hao, L.~Xu, H.~Zhu, and B.~Li, ``Automated poisoning attacks and defenses in malware detection systems: An adversarial machine learning approach,'' \emph{Computers \& Security}, vol.~73, pp. 326--344, 2018.

\bibitem{kieu2019outlier}
T.~Kieu, B.~Yang, C.~Guo, and C.~S. Jensen, ``Outlier detection for time series with recurrent autoencoder ensembles.'' in \emph{International Joint Conference on Artificial Intelligence (IJCAI)}.\hskip 1em plus 0.5em minus 0.4em\relax Morgan Kaufmann, 2019, pp. 2725--2732.

\bibitem{kieu2022anomaly}
T.~Kieu, B.~Yang, C.~Guo, R.-G. Cirstea, Y.~Zhao, Y.~Song, and C.~S. Jensen, ``Anomaly detection in time series with robust variational quasi-recurrent autoencoders,'' in \emph{IEEE International Conference on Data Engineering (ICDE)}.\hskip 1em plus 0.5em minus 0.4em\relax IEEE, 2022, pp. 1342--1354.

\bibitem{tramer2018ensemble}
F.~Tram{\`e}r, D.~Boneh, A.~Kurakin, I.~Goodfellow, N.~Papernot, and P.~McDaniel, ``Ensemble adversarial training: Attacks and defenses,'' in \emph{International Conference on Learning Representations (ICLR)}.\hskip 1em plus 0.5em minus 0.4em\relax IMLS, 2018, pp. 1--20.

\bibitem{shah2021adversarial}
J.~Li, T.~Zhu, W.~Ren, and K.-K. Raymond, ``Improve individual fairness in federated learning via adversarial training,'' \emph{Computers \& Security}, pp. 1--10, 2023.

\bibitem{chen2021certifiably}
C.~Chen, B.~Kailkhura, R.~Goldhahn, and Y.~Zhou, ``Certifiably-robust federated adversarial learning via randomized smoothing,'' in \emph{IEEE International Conference on Mobile Ad Hoc and Smart Systems (MASS)}.\hskip 1em plus 0.5em minus 0.4em\relax IEEE, 2021, pp. 173--179.

\bibitem{phuong2019towards}
M.~Phuong and C.~Lampert, ``Towards understanding knowledge distillation,'' in \emph{International Conference on Machine Learning (ICML)}.\hskip 1em plus 0.5em minus 0.4em\relax PMLR, 2019, pp. 5142--5151.

\bibitem{li2019fedmd}
E.~Sannara, F.~Portet, P.~Lalanda, and V.~German, ``A federated learning aggregation algorithm for pervasive computing: Evaluation and comparison,'' in \emph{IEEE International Conference on Pervasive Computing and Communications (PerCom)}.\hskip 1em plus 0.5em minus 0.4em\relax IEEE, 2021, pp. 1--10.

\bibitem{lin2020ensemble}
T.~Lin, L.~Kong, S.~U. Stich, and M.~Jaggi, ``Ensemble distillation for robust model fusion in federated learning,'' in \emph{Advances in Neural Information Processing Systems (NeurIPS)}.\hskip 1em plus 0.5em minus 0.4em\relax MIT Press, 2020, pp. 2351--2363.

\bibitem{guerraoui2018hidden}
R.~Guerraoui, S.~Rouault \emph{et~al.}, ``The hidden vulnerability of distributed learning in byzantium,'' in \emph{International Conference on Machine Learning (ICML)}.\hskip 1em plus 0.5em minus 0.4em\relax PMLR, 2018, pp. 3521--3530.

\bibitem{pillutla2022robust}
K.~Pillutla, S.~M. Kakade, and Z.~Harchaoui, ``Robust aggregation for federated learning,'' \emph{IEEE Transactions on Signal Processing}, vol.~70, pp. 1142--1154, 2022.

\bibitem{cao2019understanding}
D.~Cao, S.~Chang, Z.~Lin, G.~Liu, and D.~Sun, ``Understanding distributed poisoning attack in federated learning,'' in \emph{IEEE International Conference on Parallel and Distributed Systems (ICPADS)}.\hskip 1em plus 0.5em minus 0.4em\relax IEEE, 2019, pp. 233--239.

\bibitem{munoz2019byzantine}
L.~Mu{\~n}oz-Gonz{\'a}lez, K.~T. Co, and E.~C. Lupu, ``Byzantine-robust federated machine learning through adaptive model averaging,'' \emph{arXiv preprint arXiv:1909.05125}, 2019.

\bibitem{yang2019byzantine}
H.~Yang, X.~Zhang, M.~Fang, and J.~Liu, ``Byzantine-resilient stochastic gradient descent for distributed learning: A lipschitz-inspired coordinate-wise median approach,'' in \emph{2019 IEEE 58th Conference on Decision and Control (CDC)}.\hskip 1em plus 0.5em minus 0.4em\relax IEEE, 2019, pp. 5832--5837.

\bibitem{xie2018generalized}
C.~Xie, O.~Koyejo, and I.~Gupta, ``Generalized byzantine-tolerant sgd,'' \emph{arXiv preprint arXiv:1802.10116}, 2018.

\bibitem{wu2020federated}
Z.~Wu, Q.~Ling, T.~Chen, and G.~B. Giannakis, ``Federated variance-reduced stochastic gradient descent with robustness to byzantine attacks,'' \emph{IEEE Transactions on Signal Processing}, vol.~68, pp. 4583--4596, 2020.

\bibitem{wang2020model}
Y.~Wang, T.~Zhu, W.~Chang, S.~Shen, and W.~Ren, ``Model poisoning defense on federated learning: A validation based approach,'' in \emph{International Conference on Network and System Security}.\hskip 1em plus 0.5em minus 0.4em\relax Springer, 2020, pp. 207--223.

\bibitem{park2021sageflow}
J.~Park, D.-J. Han, M.~Choi, and J.~Moon, ``Sageflow: Robust federated learning against both stragglers and adversaries,'' in \emph{Advances in Neural Information Processing Systems (NeurIPS)}.\hskip 1em plus 0.5em minus 0.4em\relax MIT Press, 2021, pp. 840--851.

\bibitem{xie2019zeno}
C.~Xie, S.~Koyejo, and I.~Gupta, ``Zeno: Distributed stochastic gradient descent with suspicion-based fault-tolerance,'' in \emph{International Conference on Machine Learning (ICML)}.\hskip 1em plus 0.5em minus 0.4em\relax PMLR, 2019, pp. 6893--6901.

\bibitem{cao2019distributed}
X.~Cao and L.~Lai, ``Distributed gradient descent algorithm robust to an arbitrary number of byzantine attackers,'' \emph{IEEE Transactions on Signal Processing}, vol.~67, no.~22, pp. 5850--5864, 2019.

\bibitem{zhao2020shielding}
L.~Zhao, S.~Hu, Q.~Wang, J.~Jiang, C.~Shen, X.~Luo, and P.~Hu, ``Shielding collaborative learning: Mitigating poisoning attacks through client-side detection,'' \emph{IEEE Transactions on Dependable and Secure Computing}, vol.~18, no.~5, pp. 2029--2041, 2020.

\bibitem{zhao2020pdgan}
Y.~Zhao, J.~Chen, J.~Zhang, D.~Wu, J.~Teng, and S.~Yu, ``Pdgan: A novel poisoning defense method in federated learning using generative adversarial network,'' in \emph{International Conference on Algorithms and Architectures for Parallel Processing (ICA3PP)}.\hskip 1em plus 0.5em minus 0.4em\relax Springer, 2020, pp. 595--609.

\bibitem{li2023honest}
Y.~Li, H.~Chen, W.~Bao, Z.~Xu, and D.~Yuan, ``Honest score client selection scheme: Preventing federated learning label flipping attacks in non-iid scenarios,'' \emph{arXiv preprint arXiv:2311.05826}, 2023.

\bibitem{sattler2020byzantine}
F.~Sattler, K.-R. M{\"u}ller, T.~Wiegand, and W.~Samek, ``On the byzantine robustness of clustered federated learning,'' in \emph{International Conference on Acoustics, Speech and Signal Processing (ICASSP)}.\hskip 1em plus 0.5em minus 0.4em\relax IEEE, 2020, pp. 8861--8865.

\bibitem{ghosh2019robust}
A.~Ghosh, J.~Hong, D.~Yin, and K.~Ramchandran, ``Robust federated learning in a heterogeneous environment,'' in \emph{ICML (Workshop)}, 2019.

\bibitem{qi2024towards}
T.~Qi, H.~Wang, and Y.~Huang, ``Towards the robustness of differentially private federated learning,'' in \emph{Proceedings of the AAAI Conference on Artificial Intelligence}, vol.~38, no.~18, 2024, pp. 19\,911--19\,919.

\bibitem{singh2023fair}
A.~K. Singh, A.~Blanco-Justicia, and J.~Domingo-Ferrer, ``Fair detection of poisoning attacks in federated learning on non-iid data,'' \emph{Data Mining and Knowledge Discovery}, vol.~37, no.~5, pp. 1998--2023, 2023.

\bibitem{chen2020zero}
Z.~Chen, P.~Tian, W.~Liao, and W.~Yu, ``Zero knowledge clustering based adversarial mitigation in heterogeneous federated learning,'' \emph{IEEE Transactions on Network Science and Engineering}, vol.~8, no.~2, pp. 1070--1083, 2020.

\bibitem{li2019rsa}
L.~Li, W.~Xu, T.~Chen, G.~B. Giannakis, and Q.~Ling, ``Rsa: Byzantine-robust stochastic aggregation methods for distributed learning from heterogeneous datasets,'' in \emph{AAAI Conference on Artificial Intelligence (AAAI)}.\hskip 1em plus 0.5em minus 0.4em\relax AAAI, 2019, pp. 1544--1551.

\bibitem{zhao2024huber}
P.~Zhao, F.~Yu, and Z.~Wan, ``A huber loss minimization approach to byzantine robust federated learning,'' in \emph{Proceedings of the AAAI Conference on Artificial Intelligence}, vol.~38, no.~19, 2024, pp. 21\,806--21\,814.

\bibitem{liu2018fine}
K.~Liu, B.~Dolan-Gavitt, and S.~Garg, ``Fine-pruning: Defending against backdooring attacks on deep neural networks,'' in \emph{International Symposium on Research in Attacks, Intrusions, and Defenses (RAID)}.\hskip 1em plus 0.5em minus 0.4em\relax Springer, 2018, pp. 273--294.

\bibitem{jiang2022model}
Y.~Jiang, S.~Wang, V.~Valls, B.~J. Ko, W.-H. Lee, K.~K. Leung, and L.~Tassiulas, ``Model pruning enables efficient federated learning on edge devices,'' \emph{IEEE Transactions on Neural Networks and Learning Systems}, vol.~1, no.~1, pp. 1--13, 2022.

\bibitem{van1998introduction}
J.~H. Van~Lint, \emph{Introduction to coding theory}.\hskip 1em plus 0.5em minus 0.4em\relax Springer Science \& Business Media, 1998.

\bibitem{chen2018draco}
L.~Chen, H.~Wang, Z.~Charles, and D.~Papailiopoulos, ``Draco: Byzantine-resilient distributed training via redundant gradients,'' in \emph{International Conference on Machine Learning (ICML)}.\hskip 1em plus 0.5em minus 0.4em\relax PMLR, 2018, pp. 903--912.

\bibitem{rajput2019detox}
S.~Rajput, H.~Wang, Z.~Charles, and D.~Papailiopoulos, ``Detox: A redundancy-based framework for faster and more robust gradient aggregation,'' in \emph{Advances in Neural Information Processing Systems (NeurIPS)}.\hskip 1em plus 0.5em minus 0.4em\relax MIT Press, 2019, pp. 1--11.

\bibitem{riegercrowdguard}
P.~Rieger, T.~Krau{\ss}, M.~Miettinen, A.~Dmitrienko, and A.-R. Sadeghi, ``Crowdguard: Federated backdoor detection in federated learning,'' in \emph{Network and Distributed System Security (NDSS) Symposium}, 2024, pp. 1--18.

\bibitem{moreno2012unifying}
J.~G. Moreno-Torres, T.~Raeder, R.~Alaiz-Rodr{\'\i}guez, N.~V. Chawla, and F.~Herrera, ``A unifying view on dataset shift in classification,'' \emph{Pattern Recognition}, vol.~45, no.~1, pp. 521--530, 2012.

\bibitem{xie2020zeno++}
C.~Xie, S.~Koyejo, and I.~Gupta, ``Zeno++: Robust fully asynchronous sgd,'' in \emph{International Conference on Machine Learning (ICML)}.\hskip 1em plus 0.5em minus 0.4em\relax PMLR, 2020, pp. 10\,495--10\,503.

\bibitem{sattler2020clustered}
F.~Sattler, K.-R. M{\"u}ller, and W.~Samek, ``Clustered federated learning: Model-agnostic distributed multitask optimization under privacy constraints,'' \emph{IEEE transactions on neural networks and learning systems}, vol.~32, no.~8, pp. 3710--3722, 2020.

\bibitem{andrew2021differentially}
G.~Andrew, O.~Thakkar, B.~McMahan, and S.~Ramaswamy, ``Differentially private learning with adaptive clipping,'' in \emph{Advances in Neural Information Processing Systems (NeurIPS)}.\hskip 1em plus 0.5em minus 0.4em\relax MIT Press, 2021, pp. 17\,455--17\,466.

\bibitem{sun2021fl}
J.~Sun, A.~Li, L.~DiValentin, A.~Hassanzadeh, Y.~Chen, and H.~Li, ``Fl-wbc: Enhancing robustness against model poisoning attacks in federated learning from a client perspective,'' in \emph{Advances in Neural Information Processing Systems (NeurIPS)}.\hskip 1em plus 0.5em minus 0.4em\relax MIT Press, 2021, pp. 12\,613--12\,624.

\bibitem{gao2023sverifl}
H.~Gao, N.~He, and T.~Gao, ``Sverifl: Successive verifiable federated learning with privacy-preserving,'' \emph{Information Sciences}, vol. 622, pp. 98--114, 2023.

\bibitem{nguyen2021federated}
D.~C. Nguyen, M.~Ding, Q.-V. Pham, P.~N. Pathirana, L.~B. Le, A.~Seneviratne, J.~Li, D.~Niyato, and H.~V. Poor, ``Federated learning meets blockchain in edge computing: Opportunities and challenges,'' \emph{IEEE Internet of Things Journal}, vol.~8, no.~16, pp. 12\,806--12\,825, 2021.

\bibitem{qu2022blockchain}
Y.~Qu, M.~P. Uddin, C.~Gan, Y.~Xiang, L.~Gao, and J.~Yearwood, ``Blockchain-enabled federated learning: A survey,'' \emph{ACM Computing Surveys}, vol.~55, no.~4, pp. 1--35, 2022.

\bibitem{kim2019blockchained}
H.~Kim, J.~Park, M.~Bennis, and S.-L. Kim, ``Blockchained on-device federated learning,'' \emph{IEEE Communications Letters}, vol.~24, no.~6, pp. 1279--1283, 2019.

\bibitem{9399813}
L.~Feng, Y.~Zhao, S.~Guo, X.~Qiu, W.~Li, and P.~Yu, ``Bafl: A blockchain-based asynchronous federated learning framework,'' \emph{IEEE Transactions on Computers}, vol.~71, no.~5, pp. 1092--1103, 2022.

\bibitem{liang2018detecting}
B.~Liang, H.~Li, M.~Su, X.~Li, W.~Shi, and X.~Wang, ``Detecting adversarial image examples in deep neural networks with adaptive noise reduction,'' \emph{IEEE Transactions on Dependable and Secure Computing}, vol.~18, no.~1, pp. 72--85, 2018.

\bibitem{ross2018improving}
A.~Ross and F.~Doshi-Velez, ``Improving the adversarial robustness and interpretability of deep neural networks by regularizing their input gradients,'' in \emph{AAAI Conference on Artificial Intelligence (AAAI)}.\hskip 1em plus 0.5em minus 0.4em\relax AAAI, 2018, pp. 1--10.

\bibitem{xie2023unraveling}
C.~Xie, Y.~Long, P.-Y. Chen, Q.~Li, S.~Koyejo, and B.~Li, ``Unraveling the connections between privacy and certified robustness in federated learning against poisoning attacks,'' in \emph{Proceedings of the 2023 ACM SIGSAC Conference on Computer and Communications Security}, 2023, pp. 1511--1525.

\bibitem{zhuang2014towards}
R.~Zhuang, S.~A. DeLoach, and X.~Ou, ``Towards a theory of moving target defense,'' in \emph{ACM Workshop on Moving Target Defense (MTD)}.\hskip 1em plus 0.5em minus 0.4em\relax ACM, 2014, pp. 31--40.

\bibitem{dwork2014algorithmic}
C.~Dwork, A.~Roth \emph{et~al.}, ``The algorithmic foundations of differential privacy,'' \emph{Foundations and Trends{\textregistered} in Theoretical Computer Science}, vol.~9, no. 3--4, pp. 211--407, 2014.

\bibitem{geyer2017differentially}
R.~C. Geyer, T.~Klein, and M.~Nabi, ``Differentially private federated learning: A client level perspective,'' \emph{arXiv preprint arXiv:1712.07557}, 2017.

\bibitem{mcmahan2018learning}
H.~B. McMahan, D.~Ramage, K.~Talwar, and L.~Zhang, ``Learning differentially private recurrent language models,'' in \emph{International Conference on Learning Representations}, 2018.

\bibitem{foret2020sharpness}
P.~Foret, A.~Kleiner, H.~Mobahi, and B.~Neyshabur, ``Sharpness-aware minimization for efficiently improving generalization,'' \emph{arXiv preprint arXiv:2010.01412}, 2020.

\bibitem{cheng2022differentially}
A.~Cheng, P.~Wang, X.~S. Zhang, and J.~Cheng, ``Differentially private federated learning with local regularization and sparsification,'' in \emph{Proceedings of the IEEE/CVF Conference on Computer Vision and Pattern Recognition}, 2022, pp. 10\,122--10\,131.

\bibitem{xu2023learning}
Z.~Xu, M.~Collins, Y.~Wang, L.~Panait, S.~Oh, S.~Augenstein, T.~Liu, F.~Schroff, and H.~B. McMahan, ``Learning to generate image embeddings with user-level differential privacy,'' in \emph{Proceedings of the IEEE/CVF Conference on Computer Vision and Pattern Recognition}, 2023, pp. 7969--7980.

\bibitem{wang2020hybrid}
C.~Wang, J.~Liang, M.~Huang, B.~Bai, K.~Bai, and H.~Li, ``Hybrid differentially private federated learning on vertically partitioned data,'' \emph{arXiv preprint arXiv:2009.02763}, 2020.

\bibitem{9945997}
L.~Lyu, H.~Yu, X.~Ma, C.~Chen, L.~Sun, J.~Zhao, Q.~Yang, and P.~S. Yu, ``Privacy and robustness in federated learning: Attacks and defenses,'' \emph{IEEE Transactions on Neural Networks and Learning Systems}, pp. 1--21, 2022.

\bibitem{duchi2013local}
J.~C. Duchi, M.~I. Jordan, and M.~J. Wainwright, ``Local privacy and statistical minimax rates,'' in \emph{2013 IEEE 54th annual symposium on foundations of computer science}.\hskip 1em plus 0.5em minus 0.4em\relax IEEE, 2013, pp. 429--438.

\bibitem{truex2020ldp}
S.~Truex, L.~Liu, K.-H. Chow, M.~E. Gursoy, and W.~Wei, ``Ldp-fed: Federated learning with local differential privacy,'' in \emph{Proceedings of the third ACM international workshop on edge systems, analytics and networking}, 2020, pp. 61--66.

\bibitem{zhao2020local}
Y.~Zhao, J.~Zhao, M.~Yang, T.~Wang, N.~Wang, L.~Lyu, D.~Niyato, and K.-Y. Lam, ``Local differential privacy-based federated learning for internet of things,'' \emph{IEEE Internet of Things Journal}, vol.~8, no.~11, pp. 8836--8853, 2020.

\bibitem{fu2024differentially}
J.~Fu, Y.~Hong, X.~Ling, L.~Wang, X.~Ran, Z.~Sun, W.~H. Wang, Z.~Chen, and Y.~Cao, ``Differentially private federated learning: A systematic review,'' \emph{arXiv preprint arXiv:2405.08299}, 2024.

\bibitem{wang2023ppefl}
B.~Wang, Y.~Chen, H.~Jiang, and Z.~Zhao, ``Ppefl: Privacy-preserving edge federated learning with local differential privacy,'' \emph{IEEE Internet of Things Journal}, 2023.

\bibitem{naseri2020local}
M.~Naseri, J.~Hayes, and E.~De~Cristofaro, ``Local and central differential privacy for robustness and privacy in federated learning,'' in \emph{Network and Distributed Systems Security (NDSS) Symposium}, 2022, pp. 1--18.

\bibitem{huang2020dp}
X.~Huang, Y.~Ding, Z.~L. Jiang, S.~Qi, X.~Wang, and Q.~Liao, ``Dp-fl: a novel differentially private federated learning framework for the unbalanced data,'' \emph{World Wide Web}, vol.~23, pp. 2529--2545, 2020.

\bibitem{wei2021user}
K.~Wei, J.~Li, M.~Ding, C.~Ma, H.~Su, B.~Zhang, and H.~V. Poor, ``User-level privacy-preserving federated learning: Analysis and performance optimization,'' \emph{IEEE Transactions on Mobile Computing}, vol.~21, no.~9, pp. 3388--3401, 2021.

\bibitem{zheng2021federated}
Q.~Zheng, S.~Chen, Q.~Long, and W.~Su, ``Federated f-differential privacy,'' in \emph{International conference on artificial intelligence and statistics}.\hskip 1em plus 0.5em minus 0.4em\relax PMLR, 2021, pp. 2251--2259.

\bibitem{sunfederated}
L.~Sun and L.~Lyu, ``Federated model distillation with noise-free differential privacy,'' in \emph{International Joint Conference on Artificial Intelligence (IJCAI)}, 2021, pp. 1--8.

\bibitem{yang2023dynamic}
X.~Yang, W.~Huang, and M.~Ye, ``Dynamic personalized federated learning with adaptive differential privacy,'' \emph{Advances in Neural Information Processing Systems}, vol.~36, pp. 72\,181--72\,192, 2023.

\bibitem{smahi2024vfl}
A.~Smahi, H.~Li, W.~Han, A.~A. Fateh, and C.~C. Chan, ``Vfl-chain: Bulletproofing federated learning in the v2x environments,'' \emph{Future Generation Computer Systems}, 2024.

\bibitem{shi2023make}
Y.~Shi, Y.~Liu, K.~Wei, L.~Shen, X.~Wang, and D.~Tao, ``Make landscape flatter in differentially private federated learning,'' in \emph{Proceedings of the IEEE/CVF Conference on Computer Vision and Pattern Recognition}, 2023, pp. 24\,552--24\,562.

\bibitem{gu2023fl2dp}
C.~Gu, X.~Cui, X.~Zhu, and D.~Hu, ``Fl2dp: Privacy-preserving federated learning via differential privacy for artificial iot,'' \emph{IEEE Transactions on Industrial Informatics}, 2023.

\bibitem{kalra2023decentralized}
S.~Kalra, J.~Wen, J.~C. Cresswell, M.~Volkovs, and H.~R. Tizhoosh, ``Decentralized federated learning through proxy model sharing,'' \emph{Nature communications}, vol.~14, no.~1, p. 2899, 2023.

\bibitem{zhang2024idp}
J.~Zhang, H.~Zhu, F.~Wang, Y.~Zheng, Z.~Liu, and H.~Li, ``idp-fl: A fine-grained and privacy-aware federated learning framework for deep neural networks,'' \emph{Information Sciences}, p. 121035, 2024.

\bibitem{mo2021ppfl}
F.~Mo, H.~Haddadi, K.~Katevas, E.~Marin, D.~Perino, and N.~Kourtellis, ``Ppfl: privacy-preserving federated learning with trusted execution environments,'' in \emph{Proceedings of the 19th annual international conference on mobile systems, applications, and services}, 2021, pp. 94--108.

\bibitem{sameera2024privacy}
K.~Sameera, S.~Nicolazzo, M.~Arazzi, A.~Nocera, R.~R. KA, P.~Vinod, and M.~Conti, ``Privacy-preserving in blockchain-based federated learning systems,'' \emph{Computer Communications}, 2024.

\bibitem{singh2022framework}
S.~Singh, S.~Rathore, O.~Alfarraj, A.~Tolba, and B.~Yoon, ``A framework for privacy-preservation of iot healthcare data using federated learning and blockchain technology,'' \emph{Future Generation Computer Systems}, vol. 129, pp. 380--388, 2022.

\bibitem{kumar2021blockchain}
R.~Kumar, A.~A. Khan, J.~Kumar, N.~A. Golilarz, S.~Zhang, Y.~Ting, C.~Zheng, W.~Wang \emph{et~al.}, ``Blockchain-federated-learning and deep learning models for covid-19 detection using ct imaging,'' \emph{IEEE Sensors Journal}, vol.~21, no.~14, pp. 16\,301--16\,314, 2021.

\bibitem{li2020blockchain}
Y.~Li, C.~Chen, N.~Liu, H.~Huang, Z.~Zheng, and Q.~Yan, ``A blockchain-based decentralized federated learning framework with committee consensus,'' \emph{IEEE Network}, vol.~35, no.~1, pp. 234--241, 2020.

\bibitem{guduri2023blockchain}
M.~Guduri, C.~Chakraborty, M.~Margala \emph{et~al.}, ``Blockchain-based federated learning technique for privacy preservation and security of smart electronic health records,'' \emph{IEEE Transactions on Consumer Electronics}, 2023.

\bibitem{ramanan2020baffle}
P.~Ramanan and K.~Nakayama, ``Baffle: Blockchain based aggregator free federated learning,'' in \emph{2020 IEEE international conference on blockchain (Blockchain)}.\hskip 1em plus 0.5em minus 0.4em\relax IEEE, 2020, pp. 72--81.

\bibitem{qu2020blockchained}
Y.~Qu, S.~R. Pokhrel, S.~Garg, L.~Gao, and Y.~Xiang, ``A blockchained federated learning framework for cognitive computing in industry 4.0 networks,'' \emph{IEEE Transactions on Industrial Informatics}, vol.~17, no.~4, pp. 2964--2973, 2020.

\bibitem{lalitha2018fully}
A.~Lalitha, S.~Shekhar, T.~Javidi, and F.~Koushanfar, ``Fully decentralized federated learning,'' in \emph{Third workshop on bayesian deep learning (NeurIPS)}, vol.~2, 2018.

\bibitem{kumar2024revamping}
K.~N. Kumar, R.~Mitra, and C.~K. Mohan, ``Revamping federated learning security from a defender's perspective: A unified defense with homomorphic encrypted data space,'' in \emph{Proceedings of the IEEE/CVF Conference on Computer Vision and Pattern Recognition}, 2024, pp. 24\,387--24\,397.

\bibitem{sav2021poseidon}
S.~Sav, A.~Pyrgelis, J.~R. Troncoso-Pastoriza, D.~Froelicher, J.-P. Bossuat, J.~S. Sousa, and J.-P. Hubaux, ``Poseidon: Privacy-preserving federated neural network learning,'' in \emph{Proceedings 2021 Network and Distributed System Security Symposium}.\hskip 1em plus 0.5em minus 0.4em\relax Internet Society, 2021.

\bibitem{zhou20222d}
X.~Zhou, W.~Liang, J.~Ma, Z.~Yan, I.~Kevin, and K.~Wang, ``2d federated learning for personalized human activity recognition in cyber-physical-social systems,'' \emph{IEEE Transactions on Network Science and Engineering}, vol.~9, no.~6, pp. 3934--3944, 2022.

\bibitem{zhang2020batchcrypt}
C.~Zhang, S.~Li, J.~Xia, W.~Wang, F.~Yan, and Y.~Liu, ``$\{$BatchCrypt$\}$: Efficient homomorphic encryption for $\{$Cross-Silo$\}$ federated learning,'' in \emph{2020 USENIX annual technical conference (USENIX ATC 20)}, 2020, pp. 493--506.

\bibitem{phong2017privacy}
L.~T. Phong, Y.~Aono, T.~Hayashi, L.~Wang, and S.~Moriai, ``Privacy-preserving deep learning: Revisited and enhanced,'' in \emph{Applications and Techniques in Information Security: 8th International Conference, ATIS 2017, Auckland, New Zealand, July 6--7, 2017, Proceedings}.\hskip 1em plus 0.5em minus 0.4em\relax Springer, 2017, pp. 100--110.

\bibitem{hao2019towards}
M.~Hao, H.~Li, G.~Xu, S.~Liu, and H.~Yang, ``Towards efficient and privacy-preserving federated deep learning,'' in \emph{ICC 2019-2019 IEEE international conference on communications (ICC)}.\hskip 1em plus 0.5em minus 0.4em\relax IEEE, 2019, pp. 1--6.

\bibitem{chai2020secure}
D.~Chai, L.~Wang, K.~Chen, and Q.~Yang, ``Secure federated matrix factorization,'' \emph{IEEE Intelligent Systems}, vol.~36, no.~5, pp. 11--20, 2020.

\bibitem{fang2021privacy}
C.~Fang, Y.~Guo, Y.~Hu, B.~Ma, L.~Feng, and A.~Yin, ``Privacy-preserving and communication-efficient federated learning in internet of things,'' \emph{Computers \& Security}, vol. 103, p. 102199, 2021.

\bibitem{liu2019secure}
C.~Liu, S.~Chakraborty, and D.~Verma, ``Secure model fusion for distributed learning using partial homomorphic encryption,'' \emph{Policy-Based Autonomic Data Governance}, pp. 154--179, 2019.

\bibitem{jiang2021flashe}
Z.~Jiang, W.~Wang, and Y.~Liu, ``Flashe: Additively symmetric homomorphic encryption for cross-silo federated learning,'' \emph{arXiv preprint arXiv:2109.00675}, 2021.

\bibitem{rieyan2024advanced}
S.~A. Rieyan, M.~R.~K. News, A.~M. Rahman, S.~A. Khan, S.~T.~J. Zaarif, M.~G.~R. Alam, M.~M. Hassan, M.~Ianni, and G.~Fortino, ``An advanced data fabric architecture leveraging homomorphic encryption and federated learning,'' \emph{Information Fusion}, vol. 102, p. 102004, 2024.

\bibitem{han2023adaptive}
J.~Han and L.~Yan, ``Adaptive batch homomorphic encryption for joint federated learning in cross-device scenarios,'' \emph{IEEE Internet of Things Journal}, 2023.

\bibitem{hijazi2023secure}
N.~M. Hijazi, M.~Aloqaily, M.~Guizani, B.~Ouni, and F.~Karray, ``Secure federated learning with fully homomorphic encryption for iot communications,'' \emph{IEEE Internet of Things Journal}, 2023.

\bibitem{tian2024lattice}
H.~Tian, Y.~Wen, F.~Zhang, Y.~Shao, and B.~Li, ``Lattice based distributed threshold additive homomorphic encryption with application in federated learning,'' \emph{Computer Standards \& Interfaces}, vol.~87, p. 103765, 2024.

\bibitem{zhou2015ppdm}
J.~Zhou, Z.~Cao, X.~Dong, and X.~Lin, ``Ppdm: A privacy-preserving protocol for cloud-assisted e-healthcare systems,'' \emph{IEEE Journal of Selected Topics in Signal Processing}, vol.~9, no.~7, pp. 1332--1344, 2015.

\bibitem{hao2019efficient}
M.~Hao, H.~Li, X.~Luo, G.~Xu, H.~Yang, and S.~Liu, ``Efficient and privacy-enhanced federated learning for industrial artificial intelligence,'' \emph{IEEE Transactions on Industrial Informatics}, vol.~16, no.~10, pp. 6532--6542, 2020.

\bibitem{hu2024maskcrypt}
C.~Hu and B.~Li, ``Maskcrypt: Federated learning with selective homomorphic encryption,'' \emph{IEEE Transactions on Dependable and Secure Computing}, 2024.

\bibitem{yao1982protocols}
A.~C. Yao, ``Protocols for secure computations,'' in \emph{23rd annual symposium on foundations of computer science (sfcs 1982)}.\hskip 1em plus 0.5em minus 0.4em\relax IEEE, 1982, pp. 160--164.

\bibitem{bonawitz2017practical}
K.~Bonawitz, V.~Ivanov, B.~Kreuter, A.~Marcedone, H.~B. McMahan, S.~Patel, D.~Ramage, A.~Segal, and K.~Seth, ``Practical secure aggregation for privacy-preserving machine learning,'' in \emph{proceedings of the 2017 ACM SIGSAC Conference on Computer and Communications Security}, 2017, pp. 1175--1191.

\bibitem{khazbak2020mlguard}
Y.~Khazbak, T.~Tan, and G.~Cao, ``Mlguard: Mitigating poisoning attacks in privacy preserving distributed collaborative learning,'' in \emph{2020 29th international conference on computer communications and networks (ICCCN)}.\hskip 1em plus 0.5em minus 0.4em\relax IEEE, 2020, pp. 1--9.

\bibitem{chen2024secure}
L.~Chen, D.~Xiao, Z.~Yu, and M.~Zhang, ``Secure and efficient federated learning via novel multi-party computation and compressed sensing,'' \emph{Information Sciences}, vol. 667, p. 120481, 2024.

\bibitem{cryptoeprint:2016/086}
\BIBentryALTinterwordspacing
V.~Costan and S.~Devadas, ``Intel {SGX} explained,'' Cryptology ePrint Archive, Paper 2016/086, 2016, \url{https://eprint.iacr.org/2016/086}. [Online]. Available: \url{https://eprint.iacr.org/2016/086}
\BIBentrySTDinterwordspacing

\bibitem{chen2020training}
Y.~Chen, F.~Luo, T.~Li, T.~Xiang, Z.~Liu, and J.~Li, ``A training-integrity privacy-preserving federated learning scheme with trusted execution environment,'' \emph{Information Sciences}, vol. 522, pp. 69--79, 2020.

\bibitem{chen2023privacy}
H.~Chen, T.~Zhu, T.~Zhang, W.~Zhou, and P.~S. Yu, ``Privacy and fairness in federated learning: on the perspective of tradeoff,'' \emph{ACM Computing Surveys}, vol.~56, no.~2, pp. 1--37, 2023.

\bibitem{wu2022federated}
L.~Wu, S.~Guo, J.~Wang, Z.~Hong, J.~Zhang, and Y.~Ding, ``Federated unlearning: Guarantee the right of clients to forget,'' \emph{IEEE Network}, vol.~36, no.~5, pp. 129--135, 2022.

\end{thebibliography}

\end{document}